\documentclass[aps,prx,groupedaddress,reprint,showpacs,superscriptaddress]{revtex4-1}
\usepackage{graphicx}
\usepackage{tabularx}
\usepackage{multirow}
\usepackage{longtable}
\usepackage{dcolumn}
\usepackage{bm}
\usepackage{amsmath}
\usepackage{amssymb}
\usepackage{longtable}
\usepackage{txfonts}
\usepackage{url}
\usepackage[colorlinks=true, linkcolor=blue, anchorcolor=blue, citecolor=blue, urlcolor=magenta]{hyperref}
\usepackage[usenames,dvipsnames]{color}
\definecolor{Gray}{gray}{0.0}
\definecolor{lightGray}{gray}{0.35}

\begin{document}
\title{
The systematic study on the stability and superconductivity of Y-Mg-H compounds under high pressure
}
\author{Peng Song}
\affiliation{School of Information Science, JAIST, Asahidai 1-1, Nomi, Ishikawa 923-1292, Japan}
\author{Zhufeng Hou}
\affiliation{State Key Laboratory of Structural Chemistry, Fujian Institute of Research on the Structure of Matter, Chinese Academy of Sciences, Fuzhou 350002, China}
\author{Pedro Baptista de Castro}
\affiliation{National Institute for Materials Science, 1-2-1 Sengen, Tsukuba, Ibaraki 305-0047, Japan}
\affiliation{University of Tsukuba, 1-1-1 Tennodai, Tsukuba, Ibaraki 305-8577, Japan}
\author{Kousuke Nakano}
\affiliation{School of Information Science, JAIST, Asahidai 1-1, Nomi, Ishikawa 923-1292, Japan}
\affiliation{International School for Advanced Studies (SISSA), Via Bonomea 265, 34136, Trieste, Italy}
\author{Kenta Hongo}
\affiliation{Research Center for Advanced Computing Infrastructure, JAIST, Asahidai 1-1, Nomi, Ishikawa 923-1292, Japan}
\author{Yoshihiko Takano}
\affiliation{National Institute for Materials Science, 1-2-1 Sengen, Tsukuba, Ibaraki 305-0047, Japan}
\affiliation{University of Tsukuba, 1-1-1 Tennodai, Tsukuba, Ibaraki 305-8577, Japan}
\author{Ryo Maezono}
\affiliation{School of Information Science, JAIST, Asahidai 1-1, Nomi, Ishikawa 923-1292, Japan}

\vspace{10mm}

\date{\today}
\begin{abstract}
Motivated by recent discovery of yttrium-based high-temperature ternary
superconducting hydrides (e.g., CaYH$_{12}$, LaYH$_{12}$, and ScYH$_{6}$), 
we have employed evolutionary algorithm and first-principles calculations to
comprehensively examine the structural stability and superconductivity of
the YMgH$_{x}$  system at high pressure.
The hydrogen content $x$ and the pressure are both important
factors in the stability of these candidate structures.
We find that the stability of hydrogen-rich materials frequently
necessitates higher pressure. For instance, the pressures to
stabilize $P4/mmm$-YMgH$_{8}$ and $Cmmm$-YMgH$_{12}$ are both more than 250 GPa.
Hydrogen-less materials, such as $I4_{1}/amd$-YMgH$_{2}$ and $P6_{3}/mmc$-YMgH$_{3}$, 
can be stable at pressures as low as 100 GPa.
In addition, we find a metastable structure for YMgH$_{6}$ with the same
space group as the $P4/mmm$-YMgH$_{8}$. A metastable sodalite-like
face-centered cubic (FCC) structure is also found in YMgH$_{12}$.
These four clathrate structures of $P4/mmm$-YMgH$_{6}$, $P4/mmm$-YMgH$_{8}$, 
$Cmmm$-YMgH$_{12}$, and $Fd\bar{3}m$-YMgH$_{12}$ is made up of
H14, H18, H24, and H24 cages, respectively, in which the H-H pair exhibits
weak covalent bonding. According to phonon calculations, $P4/mmm$-YMgH$_{6}$
and $P4/mmm$-YMgH$_{8}$ require a pressure of 300 GPa to maintain dynamic
stability, however $Cmmm$-YMgH$_{12}$ and $Fd\bar{3}m$-YMgH$_{12}$ can
maintain dynamic stability at pressures of 200 GPa and 250 GPa, respectively.
Electron-phonon coupling calculations indicate that they might be potential
high-temperature superconductors, with superconductivity intimately linked
to the H cage structure. These clathrate structures exhibit a larger H derived
electron density of states at the Fermi level and also a dense H derived phonon density of states.
These two portions can interact to generate a greater electron-phonon coupling.
Therefore, the sodalite structure $Fd\bar{3}m$-YMgH$_{12}$ has a $T_\mathrm{c}$
value of 190 K and a strong electron-phonon coupling constant of 2.18.
\end{abstract}
\maketitle

\section{Introduction}
\label{sec.intro}
Recently, with the experimental verification of room temperature superconductivity
in C-S-H compounds,~\cite{2020SNI} ternary hydrides have been playing an
increasingly important role in the search for novel high-temperature superconductors.
However, due to the diversity of ternary hydrides, the difficulty in searching for
new superconducting hydrides has increased exponentially, and only about 20
potential superconducting hydrides have been predicted.~\cite{2019SUN,2020SNI,2017MA_b,2017MA,2018LI,2019LIA_b,
2017RAH,2019LIA,2021WEI,2019XIE,2017KOK,2019SHA,2019SHA,
2020ZHA_a,2020ZHA_b,2020DIC,2020GUO,2020CUI,2020LV,2020YAN,2015MUR,2019MEN}
Compared with binary superconducting hydrides, ternary superconducting hydrides
have a higher competitive possibility to achieve the room temperature superconductivity. 
The successful prediction of room temperature superconductor Li$_{2}$MgH$_{16}$ is a good demonstration.~\cite{2019SUN}

\vspace{2mm}
Among the discovered ternary hydrides, they could be classified mainly into 
three categories. The first category is a combination of metal, non-metal, and hydrogen.
In this type of combination, theoretical calculations show the following compounds 
with superconductivity, such as YSH$_{6}$, MgGeH$_{6}$, FeSeH$_{6}$, CaBH$_{6}$, LiBH$_{2}$, and so on.~\cite{2019LIA,2017MA,2017MA_b,2017KOK,2019CHE,2019GRI,2020DIC,2020LI,2020HAO}
Interestingly, the introduction of non-metal can reduce the pressure required for stability
to a certain extent. For example, LiBH$_{8}$ was predicted to have a superconducting
transition temperature ($T_\mathrm{c}$) of 126 K at 50 GPa.~\cite{2021DIC}
Therefore, it is more promising to search for the potential low-pressure-stabilized
high-temperature superconductors in these compounds.
The second category is a combination of non-metal, non-metal, and hydrogen.
In this type of combination, the theoretically predicted superconducting hydrides
include H$_{6}$SSe, CSH$_{7}$, and so on.~\cite{2018LI,2018LIU,2020CUI,2020SUN,2019LIA,2020CHA,2018NAK}
These compounds do not show any particularly obvious structural characteristics.
In addition, their phase stability is not high enough. For example, in the CSH$_{7}$ 
reported by Sun and Cui, there may be phase decomposition relative to CH$_{4}$ 
and H$_{3}$S.~\cite{2020SUN,2020CUI} For such compounds, chemical doping 
may be a more feasible way.~\cite{2020CHA,2018NAK}
For example, in the recently reported C-S-H (287.7 K at 267 GPa), the 
highest $T_\mathrm{c}$ value (270 K) can be achieved when approximately 
0.0555 carbon is incorporated into H$_{3}$S, as predicted by the virtual 
crystal approximation (VCA) method.\cite{2020SNI,2021WAN}
The third category is combinations of metal, metal, and hydrogen.
The currently reported potential superconducting compounds 
with $T_\mathrm{c}$ being greater than 200 K mainly fall into such a combination, 
such as CaYH$_{12}$, CaMgH$_{12}$, Li$_{2}$MgH$_{16}$, and so on.~\cite{2019LIA,2019SUN,2020SUK,2017RAH}
Some of these compounds are energetically favorable to form 
the clathrate structures and also likely to have robust electronic density of states (DOS) at the Fermi level.
The structures with robust DOS tend to have a large electron-phonon 
coupling constant ($\lambda$).
Generally speaking, larger $\lambda$ is easier to show high $T_\mathrm{c}$.~\cite{2004ASH}
For example, the electron-phonon coupling constant of the room 
temperature superconductor $Fd\bar{3}m$-Li$_{2}$MgH$_{16}$ can 
reach an astonishing value of 3.35.~\cite{2019SUN}

\vspace{2mm}
Our target is to search for some compounds 
with extremely high $T_\mathrm{c}$.
Therefore, we pay more attention to the third category of ternary hydrides as mentioned above.
Among the binary hydrides discovered, YH$_{10}$ has the highest $T_\mathrm{c}$ value.~\cite{2019HEI}
We first consider the introduction of the second metal element into YH$_{x}$.
Indeed there have been many reports of potential superconducting hydrides such 
as ScYH$_{6}$, CaYH$_{12}$, LaYH$_{12}$, and YKH$_{12}$.~\cite{2019LIA,2020SEM,2021WEI,2021SON,2021SON_b}
In the previous studies of binary hydrides, it has been found that the atomic number, 
atomic mass, atomic radius, number of electrons, \textit{etc.}, have significant impact 
on the final $T_\mathrm{c}$ value.~\cite{2020FLO,2020SEM_b}
Lightweight elements usually can have a higher vibration frequency and thus can 
be helpful to realize a higher $T_\mathrm{c}$ value.~\cite{2004ASH}
The superconductivity of hydrides would greatly decrease as the number of \textit{d} 
electrons and \textit{f} electrons increase. This trend is most obvious in the metal hydrides 
of the actinides and lanthanides.~\cite{2020SEM_b}
Considering the atomic mass and the absence of \textit{d} and \textit{f} electrons, 
Be, Li, Mg, Na, $etc. $ are the most ideal candidate elements to be incorporated into YH$_x$.
Among these candidates, we previously have employed the machine learning method 
based on the Gradient Boosted Tree (GBT) algorithm to screen which compounds 
may have extremely high $T_\mathrm{c}$ values.~\cite{2021SON_b}
For the 1-1-12 composition series compounds, our machine learning predictions 
show that YMgH$_{12}$ has the highest $T_\mathrm{c}$ value of 198 K among 
those of the aforementioned four candidate elements.
In addition, among the binary hydrides of the aforementioned four candidate elements, 
MgH$_{6}$ has the $T_\mathrm{c}$ value (271 K) closest to room temperature.~\cite{2015FEN}

\vspace{2mm}
Therefore, in this rapid communication, we concentrate
our attention on introducing Mg into YH$_{x}$ to study its phase
stability and superconductivity under high pressure. In particular, 
we found that the strong competition between the phase stability and 
superconductivity of YMgH$_{6}$ and that the superconductivity of 
YMgH$_{x}$ would depend strongly on the hydrogen content and 
applied high pressure. More interestingly, we found a metastable 
phase $Fd\bar{3}m$-YMgH$_{12}$ and its superconducting 
transition temperature could be high up to 190 K at 200 GPa.\\
\section{Method}
The crystal-structure search for YMgH$_{x}$ ($x$ = 2--10, 12, 14, and 16) at 100, 200, 
300 GPa was performed using the USPEX code.~\cite{2006GLA}
Each structure undergoes 4 steps of relaxation, in which the force 
on each atom and the stress tensor were optimized using the 
VASP code~\cite{1993KRE,1994KRE,1996KRE_a,1996KRE_b} based 
on the density functional theory (DFT) within the Perdew-Burke-Ernzerhof (PBE) parameterization~\cite{1996PER} 
of the generalized gradient approximation (GGA).
By comparing the relative formation enthalpies on convex hull, 
the stability of the predicted crystal structure under different pressures is discussed.
The electron-phonon coupling spectrum and superconductivity of these stable 
structure are calculated by the density functional perturbation theory implemented 
in the Quantum-ESPRESSO code.~\cite{2009GIA,2017GIA,2020GIA}
Regarding the phonon calculations, the structures of YMgH$_x$ with small number 
of atoms were carried out using the Quantum-ESPRESSO code and the others were 
performed using the  Phonopy~\cite{2015TOG} code.
The details of computational setup in the calculations of phonon, superconductivity, 
and other properties of different crystal structures of YMgH$_x$, are given in the 
Supplemental Material (SM).

\section{Results and discussion}
\label{sec.results}
\begin{figure}[htbp]
  \begin{center}
    \includegraphics[width=\linewidth]{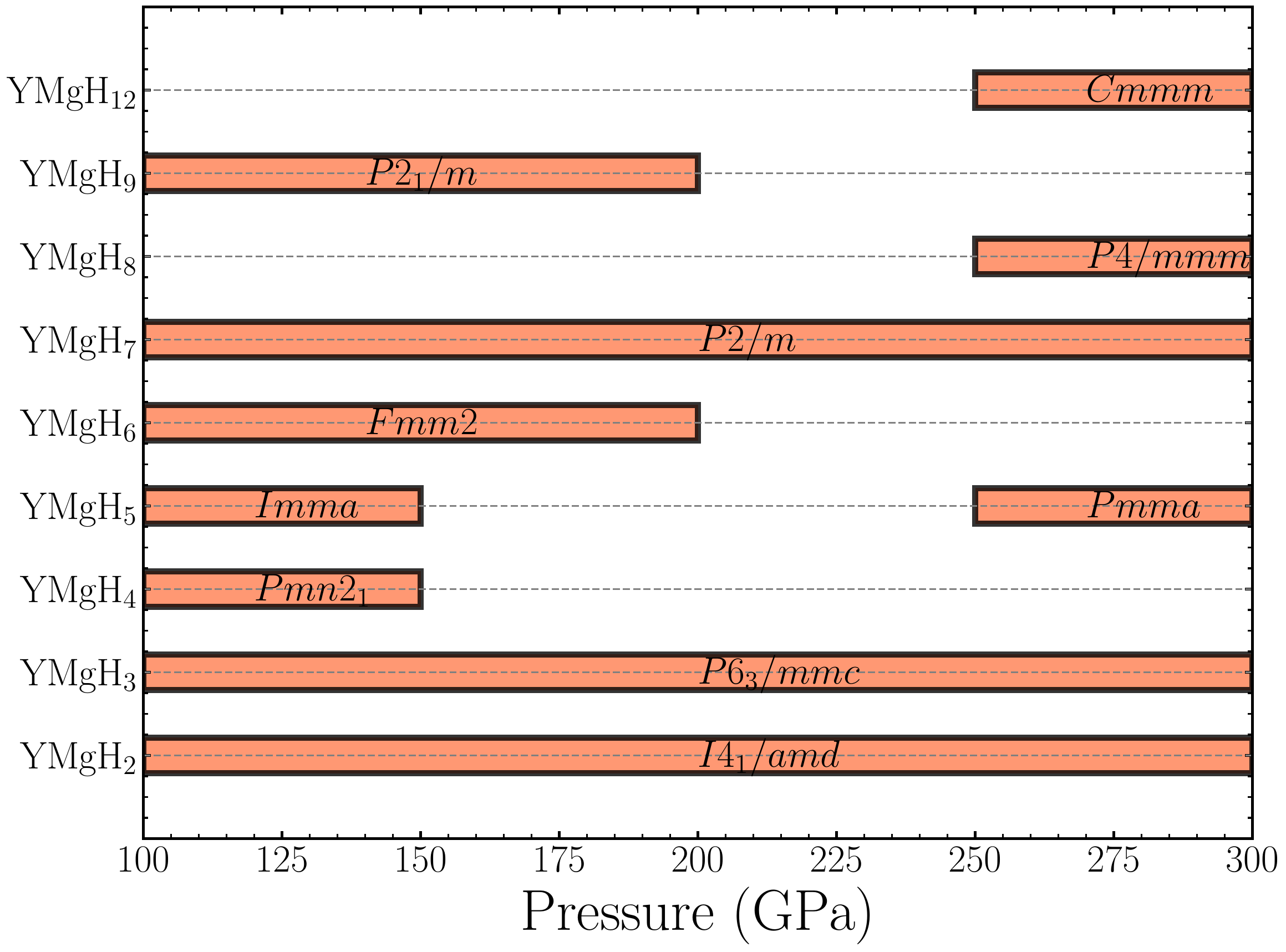}
    \caption{
Pressure-composition phase diagram of YMgH$_{x}$.
    }
    \label{fig.phase_diagram}
  \end{center}
\end{figure}

The pressure-dependent phase diagram of the thermally stabilizing YMgH$_{x}$ compounds 
in the pressure range from 100 GPa to 300 GPa is shown in Fig.~\ref{fig.phase_diagram}.
The pressure-dependent relative enthalpy and thermal convex hull of YMgH$_{x}$ are given 
in Figs.~S1 and S2 in the SM, respectively.
As shown in Fig.~S1 in the SM, $I4_{1}/amd$-YMgH$_2$ can be stabilized in a wide pressure 
range and its decomposition enthalpy indicates that it is stable 
against $P6/mmm$-YH$_{2}$~\cite{2017LIU} and $Im\bar{3}m$-Mg~\cite{2009LIU}. 
Therefore, in the thermal convex hull of YMgH$_{x}$ we choose $I4_{1}/amd$-YMgH$_{2}$ 
and $C2/c$-H$_{2}$~\cite{2007PIC} as reference.
Herein we predict 10 new stable phases of YMgH$_{x}$ ($I4_{1}/amd$-YMgH$_{2}$, 
$P6_{3}/mmc$-YMgH$_{3}$, $Pnm2_{1}$-YMgH$_{4}$, $Imma$-YMgH$_{5}$,
$Pmma$-YMgH$_{5}$, $Fmm2$-YMgH$_{6}$, $P2/m$-YMgH$_{7}$, $P4/mmm$-YMgH$_{8}$, 
$P2_{1}/m$-YMgH$_{9}$, and $Cmmm$-YMgH$_{12}$) and two metastable 
phases ($P4/mmm$-YMgH$_{6}$ and $Fd\bar{3}m$-YMgH$_{12}$).
The crystal structures of these predicted new compounds are 
depicted in Fig.~S3 in the SM. Their lattice dynamic stabilities  
have also been checked by the calculated phonon band structures, 
which are shown in Figs.~S4 and S5 in the SM.
From the pressure-dependent phase diagram of YMgH$_{x}$, 
we find that higher pressure is typically required to stabilize the hydrogen-rich materials.
To begin with, $I4_{1}/amd$-YMgH$_{2}$, $P6_{3}/mmc$-YMgH$_{3}$, 
$P2/m$-YMgH$_{7}$ all remain thermodynamically stable in the pressure 
region from 100 GPa to 300 GPa.
As seen from the phonon band structures (Figs. S3a, S3b, S4a, and S4b in the SM) 
of these three materials, we find that $I4_{1}/amd$-YMgH$_{2}$ and $P6_{3}/mmc$-YMgH$_{3}$ 
do not have imaginary frequencies within the studied pressure range, while $P2/m$-YMgH$_{7}$ 
shows imaginary frequencies from 100 GPa and these imaginary frequencies 
do not disappear with the increase of pressure.
$Pnm2_{1}$-YMgH$_{4}$ is unstable in the convex hull for the pressure of above 150 GPa.
Meanwhile, the calculated relative enthalpy indicates that $Pnm2_{1}$-YMgH$_{4}$ at 190 
GPa may undergo the phase decomposition of $P6/mmm$-YH$_{2}$~\cite{2017LIU} + 
$P6_{3}/mmc$-MgH$_{2}$~\cite{2006VAJ}.
Phonon calculations show that $Pnm2_{1}$-YMgH$_{4}$ has 
imaginary frequencies at 100 GPa, 200 GPa, and 300 GPa, however, 
these imaginary frequencies are gradually suppressed as the pressure increases.
YMgH$_{5}$ should undergo a phase transition from $Imma$ to $Pmma$ around 175 GPa.
For $Fmm2$-YMgH$_{6}$ and $P4/mmm$-YMgH$_{8}$, they could be stable in the 
pressure range of 100-150 GPa and 250-300 GPa, respectively.
In addition, we also found a metastable structure-$P4/mmm$ in YMgH$_{6}$.
The enthalpy difference between the metastable $P4/mmm$ phase and the stable 
$Fmm2$ phase of YMgH$_{6}$ at 100 GPa is 0.005 eV/atom.
The $P4/mmm$ structure was also found in ScYH$_{6}$ and ScCaH$_{8}$.~\cite{2021WEI, 2021SHI} 
More interestingly, the $T_\mathrm{c}$ value of $P4/mmm$-ScCaH$_{8}$ was predicted to be 
incredibly high up to 212 K.~\cite{2021SHI}
Regarding this context, we are curious to know whether $P4/mmm$-YMgH$_{6}$ and 
$P4/mmm$-YMgH$_{8}$ would exhibit superconductivity or not, since they possess 
the same space group as ScCaH$_{8}$.
At 200 GPa the phonons of $P4/mmm$-YMgH$_{6}$ and $P4/mmm$-YMgH$_{8}$ 
show imaginary frequencies, however, at 300 GPa these two phases attain the lattice dynamic stability.
In the thermal convex hull diagram, $P2_{1}/m$-YMgH$_{9}$ could be thermodynamically stabilized 
in the pressure range of 100-200 GPa, however, its phonons always have imaginary 
frequencies in such a pressure range. Therefore, the lattice dynamic of $P2_{1}/m$-YMgH$_{9}$ is unstable.
We predicted two highly symmetrical clathrate structures for YMgH$_{12}$, i.e, the 
$Cmmm$ phase and the metastable $Fd\bar{3}m$ phase. Their crystal structures 
are depicted in Fig.~\ref{fig.elf_1}. These two clathrate structures are composed 
of H$_{24}$ cages, each with six quadrilaterals and eight hexagons.
According to the binary hydrides of Y-H and Mg-H, we analyzed the probable 
decomposition path of YH$_{6}$($Im\bar{3}m$)~\cite{2017LIU} + 
MgH$_{2}$($P6_{3}/mmc$)~\cite{2006VAJ} + 2H$_{2}$($C2/c$)~\cite{2007PIC} 
for YMgH$_{12}$. The calculated results indicate that $Cmmm$-YMgH$_{12}$ 
can also be stable over 250 GPa.
The calculated phonon dispersion suggests that the lattice dynamics of both 
the $Cmmm$ and $Fd\bar{3}m$ phases of YMgH$_{12}$ can stay stable over 
250 GPa. We should point out that the phases of YMgH$_x$ with $x$ = 10, 14, and 16 
in the studied pressure range are above the thermodynamic convex hull. In the 
proceeding subsections we will focus on the thermodynamically stable and 
metastable phases of YMgH$_x$ with high lattice dynamic stability.

\begin{figure}[htbp]
  \begin{center}
    \includegraphics[width=\linewidth]{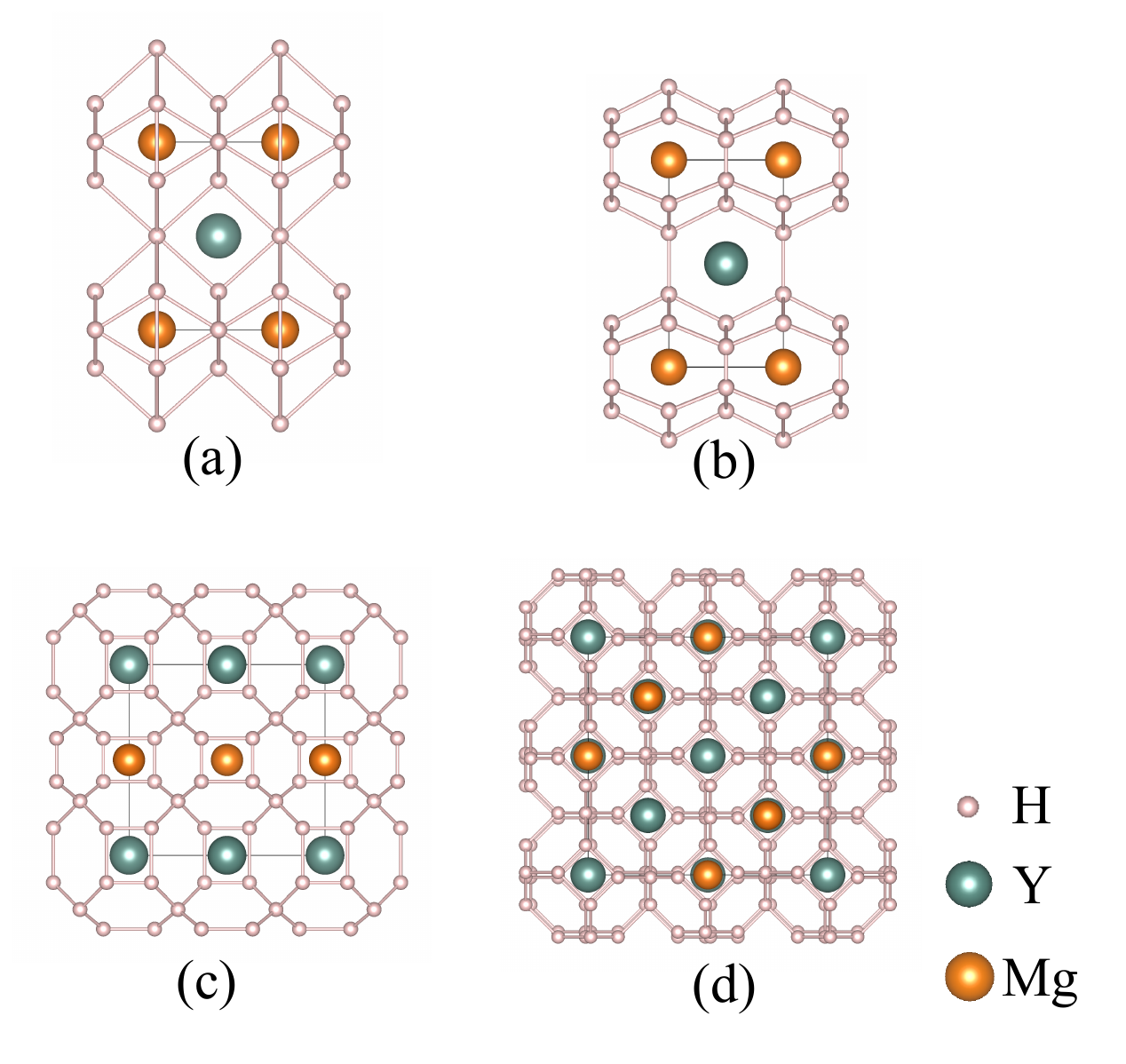}
    \caption{The clathrate structure of (a) $P4/mmm$-YMgH$_{6}$,  (b) $P4/mmm$-YMgH$_{8}$,  (c) $Cmmm$-YMgH$_{12}$,  (d) $Fd\bar{3}m$-YMgH$_{12}$.}
    \label{fig.elf_1}
  \end{center}
\end{figure}

\begin{figure}[htb]
  \begin{center}
    \includegraphics[width=\linewidth]{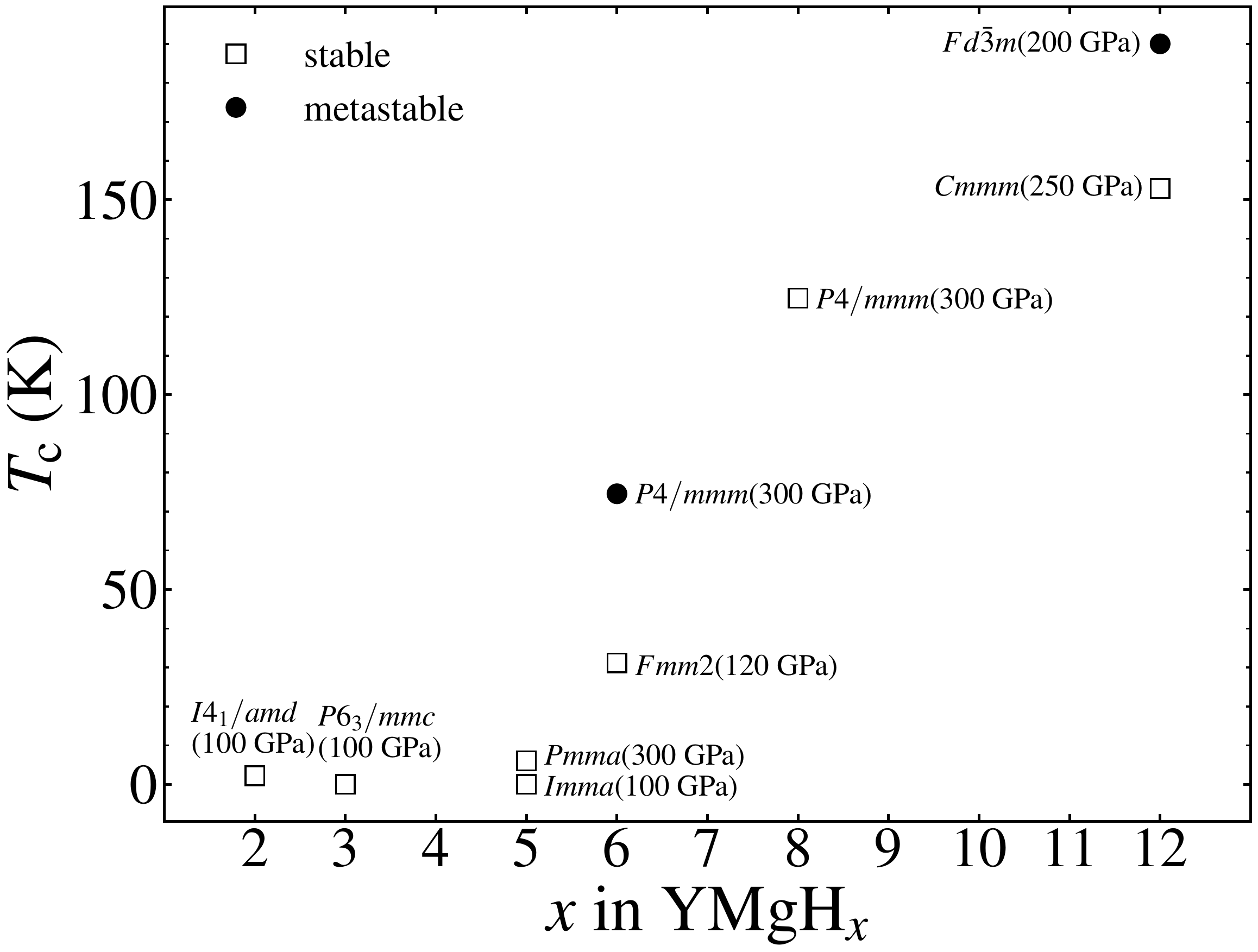}
    \caption{
Calculated superconducting temperature $T_\mathrm{c}$ ( $\mu$ = 0.1) of various structures at high pressure.
    }
    \label{fig.Tc_H}
  \end{center}
\end{figure}

To discover the superconductivity in YMgH$_{x}$, we have calculated the electron-phonon 
coupling (EPC) spectra of the stable and metastable phases of YMgH$_{x}$ ($x$ = 2, 3, 5, 6, 8, and 12) 
at certain pressures. The obtained Eliashberg spectral functions of YMgH$_{x}$ are presented 
in Fig.~S6 in the SM. The predicted $T_\mathrm{c}$ values of YMgH$_{x}$ are summarized 
in Fig.~\ref{fig.Tc_H} and also tabulated in Table~S2 in the SM. 
We find that the $I4_{1}/amd$-YMgH$_{2}$, $P6_{3}/mmc$-YMgH$_{2}$, $Imma$-YMgH$_5$ can exhibit superconductivity at high pressure, however, their $T_\mathrm{c}$ values at the pressure of 
100 GPa are smaller than 2.5 K. For YMgH$_5$, the $Pmma$ phase can be stabilized by increasing 
the pressure ($\geq$ 250 GPa) and the corresponding $T_\mathrm{c}$ value is also increased 
slightly. In particular, the YMgH$_{x}$ with $x$ = 6, 8, and 12 at high pressure exhibit 
superconductivity with much higher $T_\mathrm{c}$. From the stable $Fmm2$ phase of 
YMgH$_6$ to the stable $Cmmm$ phase of YMgH$_{12}$, the $T_\mathrm{c}$ value 
increases from $\sim$31 K (at 120 GPa) to $\sim$153 K (at 250 GPa). We also find that 
the metastable $Fd\bar{3}m$ phase of YMgH$_{12}$ could have a much higher 
$T_\mathrm{c}$ up to about 189 K (at 300 GPa). Therefore, the $T_\mathrm{c}$ value of 
YMgH$_{x}$ can be boosted by increasing the H content and the pressure.

To understand the trend of the superconductivity in YMgH$_{x}$, we have made a close examination 
of the phonon spectra and electronic structures of YMgH$_{2}$, YMgH$_{4}$, YMgH$_{5}$, YMgH$_{6}$, 
YMgH$_{8}$, and YMgH$_{12}$. The corresponding results are presented in Figs.~S4, S5, and~S7 in the SM. 
Because of the different mass among Y, Mg, and H atoms, the phonons with lower frequencies ($\lesssim$ 15 THz) 
are dominated by the vibrations of Y and Mg atoms, whereas the high-frequency phonons ($\gtrsim$ 15 THz) is
contributed by the vibration of H atoms. Generally speaking, the EPC constant $\lambda$ at 
the low-frequency region of YMgH$_{x}$ is rather small, that is to say, the contribution of Y and 
Mg atoms to the EPC is also small. From YMgH$_2$ to YMgH$_{12}$, we find that the high-frequency 
phonon density of states increase with the increase of H content $x$. For the $I41/amd$-YMgH$_{2}$, 
$P63/mmc$-YMgH$_{3}$, and $Imma$-YMgH$_{5}$, their phonon density of states at the high-frequency 
region are very small, meanwhile the electronic density of states at the Fermi level ($E_\mathrm{F}$) are also 
small and contributed dominantly by the Y-4\textit{d} orbitals, leading to a rather 
weak EPC (i.e., $\lambda \lesssim 0.4$). As a result, the $T_\mathrm{c}$ values of 
the $I41/amd$-YMgH$_{2}$, $P63/mmc$-YMgH$_{3}$, and $Imma$-YMgH$_{5}$ are very 
small accordingly. Starting with the stable $Fmm2$ phase of YMgH$_{6}$, the EPC constant 
increase dramatically (i.e., nearly doubled). However the electronic states at the $E_\mathrm{F}$ 
of the $Fmm2$-YMgH$_{6}$ is much smaller than those of the metastable $P6/mmm$-YMgH$_{6}$. 
In the latter case, the contribution of H atoms to the electronic states at the $E_\mathrm{F}$ is comparable 
to that of the Y atom. On the other hand, the metastable $P6/mmm$-YMgH$_{6}$ exhibits stronger EPC 
than the stable $Fmm2$-YMgH$_{6}$. Therefore, the metastable $P6/mmm$-YMgH$_{6}$ has a 
higher $T_\mathrm{c}$ value. For the $P4/mmm$-YMgH$_{8}$, the vibration of H atoms makes a 
small contribution to the phonons in the low-frequency range of 0-25 THz, whose dominant 
contribution comes form the vibration of Y and Mg atoms. Meanwhile the H and Y atoms have 
almost equal contribution to the electronic states at the $E_\mathrm{F}$ of the $P4/mmm$-YMgH$_{8}$. 
As a result, these features lead to a rather strong EPC (i.e., $\lambda \simeq$ 1.635) and a $T_\mathrm{c}$ 
value high up to 124 K for the $P4/mmm$-YMgH$_{8}$. The $Cmmm$ phase of YMgH$_{12}$ can be 
stabilized in the pressure from  250 GPa to 300 GPa. It is found that the phonon density of states in the 
high-frequency range of the $Cmmm$-YMgH$_{12}$ increase further as compared with the 
$P4/mmm$-YMgH$_8$, however the former one has a slightly smaller electronic states at the 
$E_\mathrm{F}$. The EPC constant of $Cmmm$-YMgH$_{12}$ is in the same order as the 
$P4/mmm$-YMgH$_8$. For the metastable $Fd\bar{3}m$-YMgH$_{12}$, the electronic states 
at the $E_\mathrm{F}$ are slightly larger than those of the stable $Cmmm$-YMgH$_{12}$ and its 
EPC is also stronger than that of the stable $Cmmm$-YMgH$_{12}$. Therefore, the 
metastable $Fd\bar{3}m$-YMgH$_{12}$ has a higher $T_\mathrm{C}$ value than the 
stable $Cmmm$-YMgH$_{12}$. In both the stable $Cmmm$ and metastable $Fd\bar{3}m$ 
phases of YMgH$_{12}$, the contribution of H atoms to the electronic states at the $E_\mathrm{F}$ 
is greater than that of the Y atom. In short, the H content and the weight of H atoms in the electronic 
states at the $E_\mathrm{F}$ play an important role in the superconductivity of YMgH$_{x}$.
\begin{figure*}[htb]
  \begin{center}
    \includegraphics[width=\linewidth]{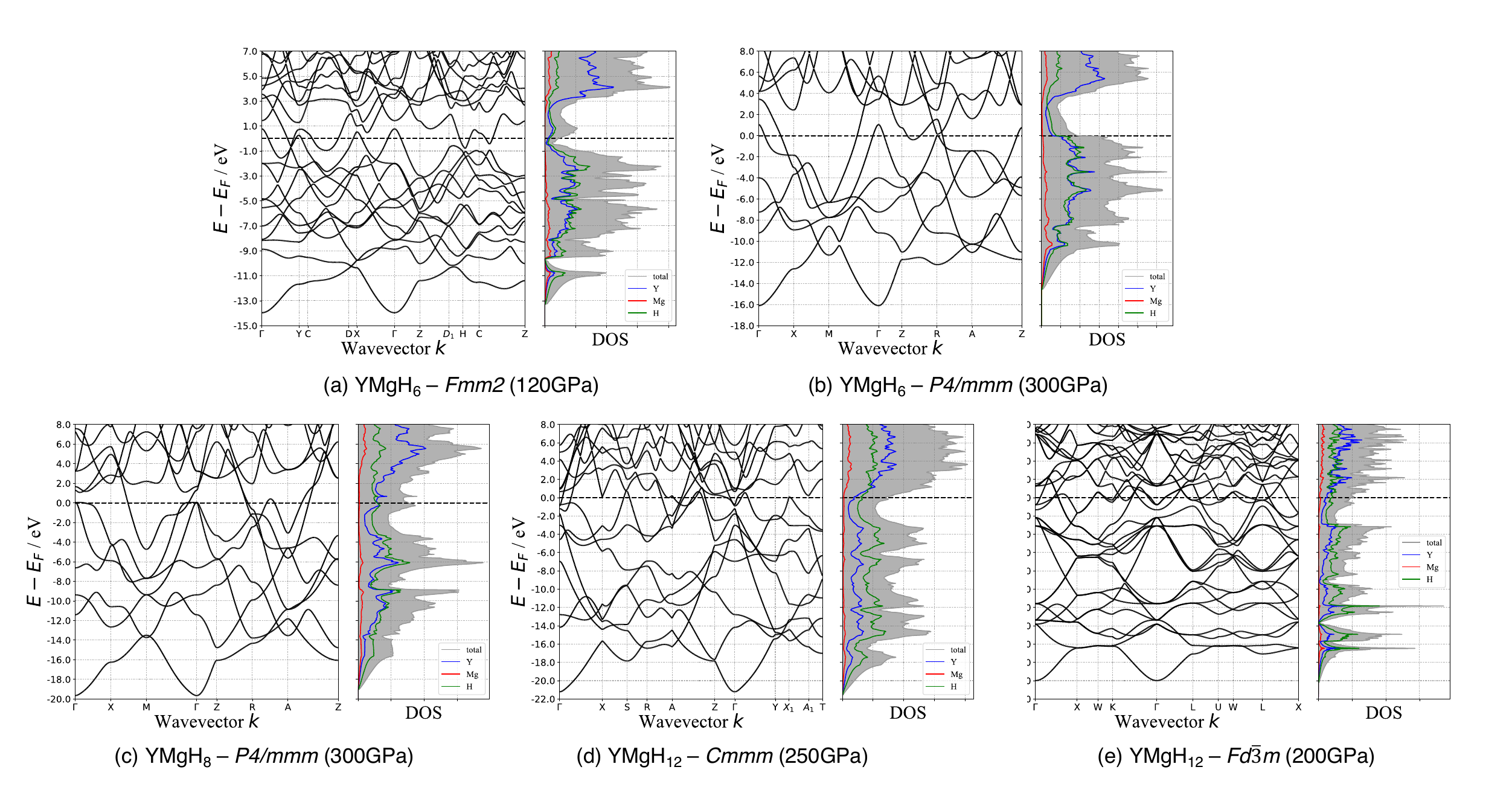}
    \caption{
Electronic band structures and density of states (DOS) for (a) $Fmm2$-YMgH$_{6}$ (120 GPa), (b) $P4/mmm$-YMgH$_{6}$ (300 GPa), (c) $P4/mmm$-YMgH$_{8}$ (300 GPa), (d) $Cmmm$-YMgH$_{12}$ (250 GPa), and (e) $Fd\bar{3}m$-YMgH$_{12}$ (200 GPa).
    }
    \label{fig.band}
  \end{center}
\end{figure*}

The intriguing thing is why the weight of H atoms in the electronic states at the 
$E_\mathrm{F}$ for the hydrogen-rich YMgH$_x$ at high pressure become more 
significant. Driven by such a question, we carried out the structure analysis and the 
further examination of the chemical bonding information of the above stable phases. 
The radial distribution functions of H-H atom pairs in YMgH$_x$ ($x$ = 2, 3, 4, 5, 6, 8, and 12) 
are presented in Fig.~S8 in the SM. We find that the first nearest-neighbour distance (1NND) 
between H atoms in YMgH$_{x}$ typically decreases as the H content increases.
The 1NND of H-H in $P4/mmm$-YMgH$_{8}$ is 0.864 \AA, which is very close to the 
H-H bond length (0.74 \AA) of hydrogen molecule.
The 1NNDs of H-H in the $Cmmm$ and $Fd\bar{3}m$ phases of YMgH$_{12}$ are 1.040 
and 1.097 \AA~respectively, which are quite close to those (i.e., 1.19 \AA~and 1.10 \AA) in 
YH$_{6}$ and MgH$_{6}$, respectively.~\cite{2017LIU,2015FEN}
This is because the formed clathrate structures of $Cmmm$-YMgH$_{12}$ 
and $Fd\bar{3}m$-YMgH$_{12}$ have similarities with MgH$_{6}$ and YH$_{6}$, 
namely, they are all composed of H$_{24}$ cages. The shorter 1NND of H-H indicates 
that the H atoms may exhibit a molecule-like state to be bonded with the Y and Mg atoms. 
Generally speaking, if some electrons fill the antibonding state of H$_2$ molecule, the H-H 
bond length would be elongated. The slightly elongated 1NNDs of H-H in $P4/mmm$-YMgH$_{8}$, 
$Cmmm$-YMgH$_{12}$, $Fd\bar{3}m$-YMgH$_{12}$ than the H-H bond length of H$_2$ 
molecule may be partially ascribed to the charge transfer from Y or Mg atoms to the H-H atom pair.

The electron localization function (ELF) of YMgH$_x$ ($x$ = 6, 8, and 12) is 
presented in Fig.~\ref{fig.elf_1}. The orbital-decomposed density of states and 
projected crystal orbital Hamilton population (pCOHP) of YMgH$_x$ are presented in 
Fig.~S9 in the SM. The shorter 1NND between H atoms also indicates that the H-H atom 
pair may form stronger chemical bonding. For instance, the 1NND of H-H in the 
$I4_{1}/amd$-YMgH$_2$ is about 2.60 \AA, which is much larger than that in 
the $P4/mmm$-YMgH$_6$. In the $I4_{1}/amd$-YMgH$_2$ (see Figs.~S3a and S9a in 
the SM), the pCOHP for the first nearest-neighbor (1NN) H-H atom pair is almost zero, 
indicating no hybridization for them. On the other hand, the H 1\textit{s} orbital makes a 
significant contribution to the valence states from -10 eV to -5 eV below the 
$E_\mathrm{F}$ of $I4_{1}/amd$-YMgH$_2$, and it is also hybridized with the 
Y 4\textit{d} orbitals and the Mg 3\textit{s} orbital to form the bonding states. 
As the H content $x$ in YMgH$_x$ increases, the energy range for the valence 
states with a significant contribution from the H-1\textit{s} orbital become wider, 
which could be associated with the change in the chemical bonding of either the
 H-H or H-Y. In the $P6_3/mmc$-YMgH$_3$, the hybridization between the 
 H-1\textit{s} orbital and the Y-4\textit{d} orbitals is greatly enhanced (see Fig.~S9b 
 in the SM), while the H-H hybridization is still almost none because the corresponding 
 1NND is about 1.98 \AA. In the $P4/mmm$-YMgH$_{8}$ (see Fig.~S9g in the SM), 
 the hybridization between the H-1\textit{s} orbital and the Y-4\textit{d} orbitals is 
 further enhanced so that the energy range for the bonding states of H-Y is extended 
 from the $E_\mathrm{F}$ to -15 eV, meanwhile the hybridization between the H-H atom 
 pair is also enhanced significantly so that the corresponding states become deeper and 
 to be centerized around -15 eV. The anti-bonding states of H-H appear around the 
 $E_\mathrm{F}$ of the $P4/mmm$-YMgH$_{8}$. In the $Cmmm$-YMgH$_{12}$ and 
 $Fd\bar{3}m$-YMgH$_{12}$ the anti-bonding states of H-H appear just below the 
 $E_\mathrm{F}$ (see Figs.~S9h and S9i in the SM). Therefore, the H-H atom pair 
 forms strong covalent bonding.  This also can be seen from the ELF, as shown in Fig.~\ref{fig.elf_1}d.
From the analysis of DOS of YMgH$_x$, we find that the contribution of H-1\textit{s} electrons to 
the electronic states at the $E_\mathrm{F}$ tends to increase as the H content $x$ increases. 
For the $I4_{1}/amd$-YMgH$_2$ and $P6_3/mmc$-YMgH$_3$, the states at the $E_\mathrm{F}$ 
are dominated by the Y-4\textit{d} orbitals and the contribution other orbital could be negligible. 
For the $Pmma$-YMgH$_5$, YMgH$_6$, YMgH$_{8}$, and YMgH$_{12}$, the contribution of the 
H atoms to the states at the $E_\mathrm{F}$ become more robust and is comparable to the contribution 
of the Y-4\textit{d} orbitals. This is due to the filling of anti-bonding states of H-H atom pair by the 
electron transferred from metal atoms.
\begin{figure*}[htbp]
  \begin{center}
    \includegraphics[width=\linewidth]{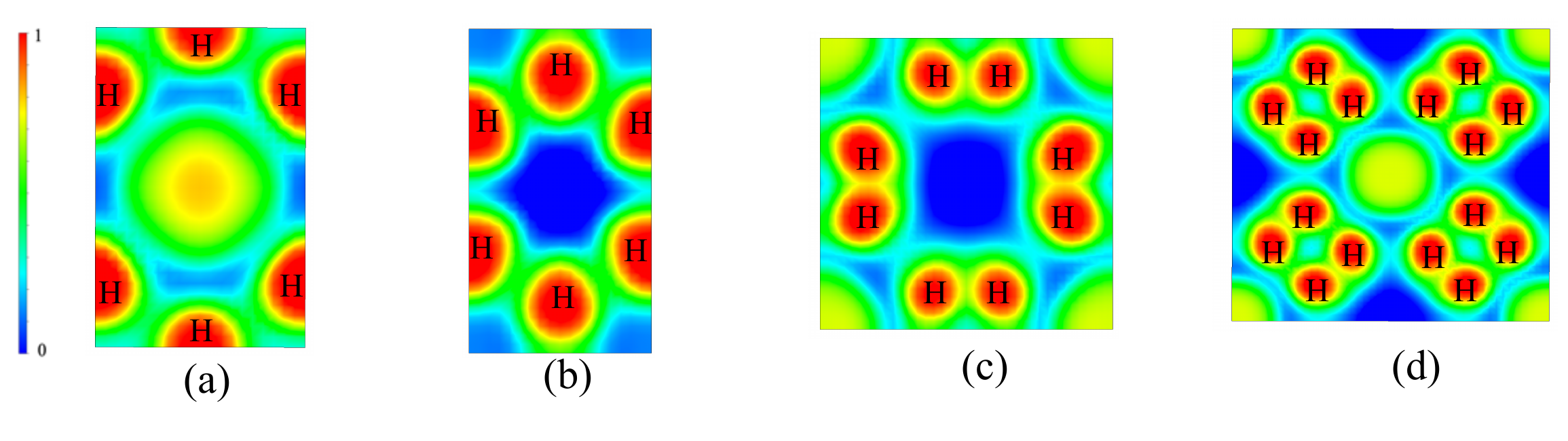}
    \caption{
The contour plots of electron localization function (ELF) on (a) (200) plane of $P4/mmm$-YMgH$_{6}$, (b) (200) plane of $P4/mmm$-YMgH$_{8}$, (c) (100) plane of $Cmmm$-YMgH$_{8}$, and (d) (100) plane of $Fd\bar{3}m$-YMgH$_{12}$
    }
    \label{fig.elf_1}
  \end{center}
\end{figure*}

\section{Conclusion}
\label{sec.conc}
In summary, we systematically investigated the stability and superconductivity of the ternary 
hydride YMgH$_{x}$ at high pressure.
For YMgH$_{x}$ with $x$ = 6, 8, and 12, they thermodynamically prefer to take the 
clathrate structure composed of H$_{14}$, H$_{18}$, and H$_{24}$ cages, respectively, 
in which strong covalent connections between H-H can be formed.
According to the EPC calculations, these clathrate structures of YMgH$_{x}$ all can 
exhibit high-temperature superconductivity.
The interplay of the extraordinarily high weight in the electronic density of states at 
the Fermi level and the dense phonon density of states associated with H atoms in these 
clathrate structures of YMgH$_{x}$ results in a significant electron-phonon coupling.
The $Fd\bar{3}m$-YMgH$_{12}$, a metastable sodalite structure, exhibits a $T_\mathrm{c}$ 
of up to 190 K at 200 GPa.
The cage structure of the superconducting ternary hydride resembles that of the 
superconducting binary hydride in a similar way.
For example, H$_{18}$ cages are also present in both $I4/mmm$-YH$_{4}$~\cite{2017LIU} 
and $P4/mmm$-YMgH$_{8}$. H$_{24}$ cages are also found in MgH$_{6}$~\cite{2015FEN}, 
$Im\bar{3}m$-YH$_{6}$~\cite{2019HEI}, $Cmmm$-YMgH$_{12}$, and $Fd\bar{3}m$-YMgH$_{12}$.
These results suggest that the search for new potential high-temperature superconducting ternary
hydrides can be substantially accelerated by starting with the structure of 
the superconducting binary clathrates.

\section{Acknowledgments}
The computations in this work have been performed
using the facilities of
Research Center for Advanced Computing
Infrastructure (RCACI) at JAIST.
K.H. is grateful for financial support from
the HPCI System Research Project (Project ID: hp190169) and
MEXT-KAKENHI (JP16H06439, JP17K17762, JP19K05029, and JP19H05169).
R.M. is grateful for financial supports from
MEXT-KAKENHI (19H04692 and 16KK0097),
FLAGSHIP2020 (project nos. hp1
90169 and hp190167 at K-computer),
Toyota Motor Corporation, I-O DATA Foundation,
the Air Force Office of Scientific Research
(AFOSR-AOARD/FA2386-17-1-4049;FA2386-19-1-4015),
and JSPS Bilateral Joint Projects (with India DST).

\bibliographystyle{apsrev4-1}
\bibliography{references}


\end{document}


\title{{\Large Supplemental Material}\\
  \vspace{3mm}
  for\\
  \vspace{3mm}
The systematic study on the stability and superconductivity of Y-Mg-H compounds under high pressure}

\author{Peng Song}
\affiliation{School of Information Science, JAIST, Asahidai 1-1, Nomi, Ishikawa 923-1292, Japan}
\author{Zhufeng Hou}
\affiliation{State Key Laboratory of Structural Chemistry, Fujian Institute of Research on the Structure of Matter, Chinese Academy of Sciences, Fuzhou 350002, China}
\author{Pedro Baptista de Castro}
\affiliation{National Institute for Materials Science, 1-2-1 Sengen, Tsukuba, Ibaraki 305-0047, Japan}
\affiliation{University of Tsukuba, 1-1-1 Tennodai, Tsukuba, Ibaraki 305-8577, Japan}
\author{Kousuke Nakano}
\affiliation{School of Information Science, JAIST, Asahidai 1-1, Nomi, Ishikawa 923-1292, Japan}
\affiliation{International School for Advanced Studies (SISSA), Via Bonomea 265, 34136, Trieste, Italy}
\author{Kenta Hongo}
\affiliation{Research Center for Advanced Computing Infrastructure, JAIST, Asahidai 1-1, Nomi, Ishikawa 923-1292, Japan}
\author{Yoshihiko Takano}
\affiliation{National Institute for Materials Science, 1-2-1 Sengen, Tsukuba, Ibaraki 305-0047, Japan}
\affiliation{University of Tsukuba, 1-1-1 Tennodai, Tsukuba, Ibaraki 305-8577, Japan}
\author{Ryo Maezono}
\affiliation{School of Information Science, JAIST, Asahidai 1-1, Nomi, Ishikawa 923-1292, Japan}


\maketitle

\subsection{Computational details}
\label{computational}
The crystal structure prediction for YMgH$_{x}$ ($x$ = 2-10, 12, 14, and 16)
at the high pressure ranging from 100 GPa to 300 GPa were carried out
using the USPEX (Universal Structure Predictor: Evolutionary
Xtallography)~\cite{2006GLA} code combined with the
VASP (Vienna \emph{ab initio} simulation
package)~\cite{1993KRE,1994KRE,1996KRE_a,1996KRE_b} code.
These chemical compositions are created at random, with the
first 400 structures being created at random, and each
subsequent generation producing 100 structures.
These structures are composed of 40\% heredity, 40\% random,
10\% mutation, and 10\% soft mutation.
The energy of the produced structures is minimized globally using
Perdew-Burke-Ernzerhof (GGA-PBE)~\cite{1996PER} through three DFT computations.
We choose around ten of the structures with the lowest structural energy
for each chemical composition and undertake two higher-precision optimizations
to guarantee that they can converge well under 1 meV.
Finally, the structure with the lowest energy is chosen as the final contender among these structures.
The LOBSTER algorithm is used to compute the output result coupled
with VASP for the crystal orbital Hamiltonian population (COHP) and
the integrated crystal orbital Hamiltonian population (ICOHP).~\cite{2016MAI,1993KRE,1994KRE,1996KRE_a,1996KRE_b}
The supercell method, as implemented in the Phonopy code,~\cite{2015TOG} was used to calculate phonons for YMgH$_4$, YMgH$_7$, and YMgH$_9$.
Quantum ESPRESSO is used to compute the phonons of
YMgH$_2$, YMgH$_3$, YMgH$_5$, YMgH$_6$, YMgH$_8$, and
YMgH$_{12}$.~\cite{2009GIA,2017GIA,2020GIA}
The electron-phonon coupling (EPC) constant $\lambda$ of these
stable compounds was calculated by quantum ESPRESSO with a plane
wave basis set up to a cutoff energy of 60 Ry and Marzari-Vanderbilt
method in the context of linear response theory.
The EPC calculation matrix is shown in Table~\ref{tab:mesh}.
The superconducting transition temperature $T_{c}$ is calculated
using McMillan's equation.~\cite{1968MCM}
\begin{equation}
T_{c} = \frac{\Theta}{1.45} \rm{exp} [ - \frac{1.04 (1 + \lambda)}{\lambda - \mu^{*} (1 + 0.62 \lambda)}]
\end{equation}
where $\Theta$ is the Debye temperature of the system. $\Theta$  = $\frac{1.45}{1.2}$ $\omega_{log}$. This $\omega_{log}$ is the logarithmic average of the spectral function, which was introduced by Dynes.
\begin{table}[htp]
\caption{\label{tab:mesh}
The setup of $k$-point mesh in the self-consistent field (SCF) calculations and of $q$-point mesh in the phonon calculations.
}
\begin{tabular}{lllllll}
\hline
Compound          & $I4_{1}/amd$-YMgH$_2$ & $P6_{3}/mmc$-YMgH$_3$ & $Imma$-YMgH$_5$ & $Pmma$-YMgH$_5$ & $Fmm2$-YMgH$_6$ & $P4/mmm$-YMgH$_6$   \\
\hline
SCF dense $k$-point & 12 $\times$ 12 $\times$ 12      & 16 $\times$ 16 $\times$ 8      & 12 $\times$ 12 $\times$ 12   & 12 $\times$ 12 $\times$ 12      & 12 $\times$ 12 $\times$ 12      & 16 $\times$ 16 $\times$ 12              \\
phonon $q$-kpoint   &  3 $\times$ 3 $\times$ 3     & 4 $\times$ 4 $\times$ 2      & 3 $\times$ 3 $\times$ 3      &  3 $\times$ 3 $\times$ 3     &  3 $\times$ 3 $\times$ 3      &  4 $\times$ 4 $\times$ 3           \\
\hline
   \specialrule{0em}{1pt}{1pt}
Compound       & $P4/mmm$-YMgH$_8$ & $Cmmm$-YMgH$_{12}$ & $Fd\bar{3}m$-YMgH$_{12}$ & &  &\\
\hline
SCF dense $k$-point & 16 $\times$ 16 $\times$ 12      & 16 $\times$ 16 $\times$ 12      &  16 $\times$ 16 $\times$ 16      &       &       &             \\
phonon $q$-kpoint   &  4 $\times$ 4 $\times$ 3     &  4 $\times$ 4 $\times$ 3     &   4 $\times$ 4 $\times$ 4    &       &       &             \\
\hline
\end{tabular}
\end{table}

\clearpage
\newpage
\subsection{ Supplementary figures}
\label{supp_figure}

\begin{figure*}[htbp]
  \begin{center}
    \includegraphics[width=\linewidth]{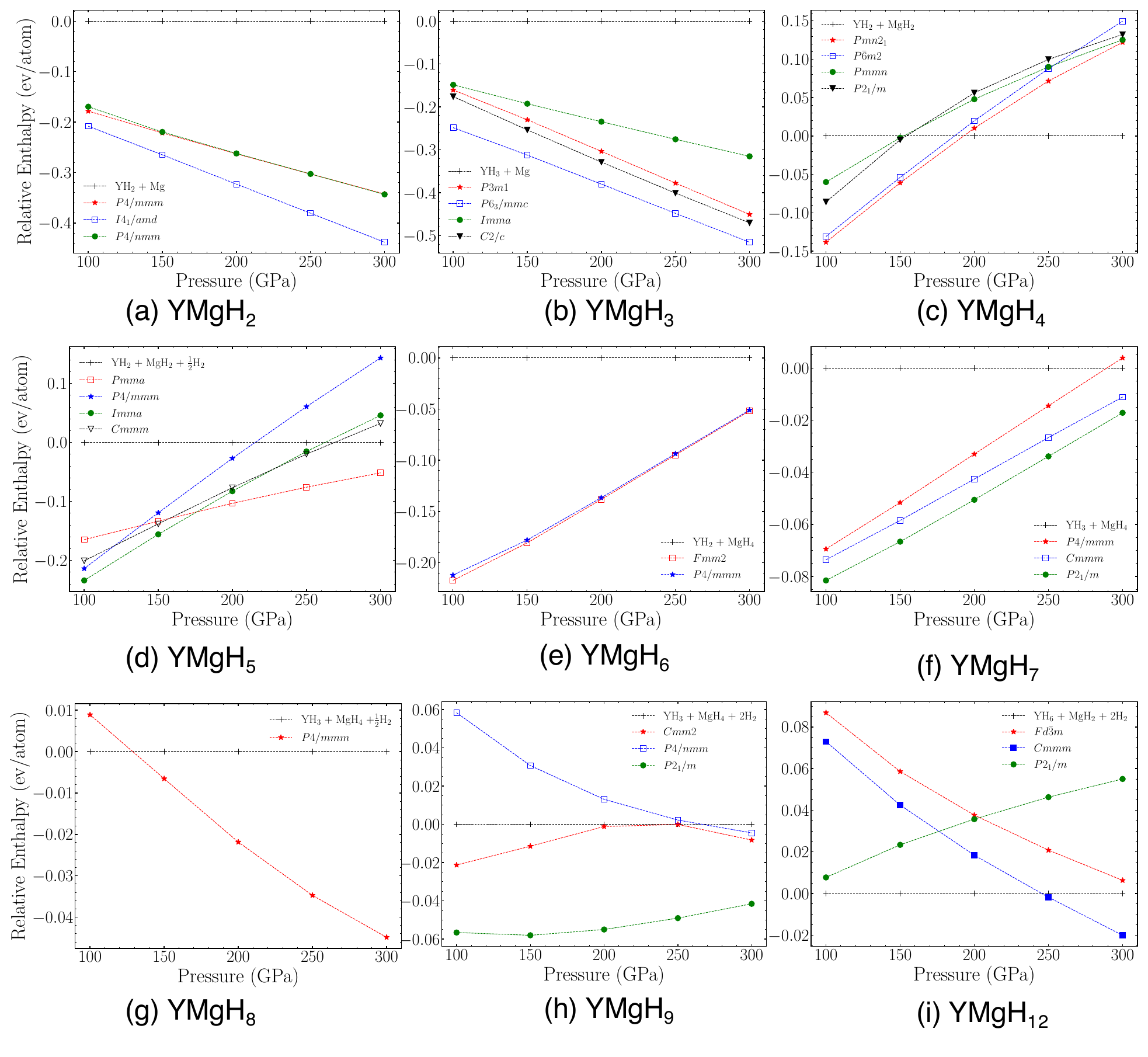}
    \caption{
Comparisons of enthalpies among the candidate structures for the YMgH$_{x}$ at the pressure ranging from 100 GPa to 300 GPa.  The decomposition enthalpy is compared with H$_{2}$,~\cite{2007PIC} Mg,~\cite{2009LIU} MgH$_{2}$,~\cite{2013LON} MgH$_{4}$,~\cite{2013LON} YH$_{2}$,~\cite{2017LIU} YH$_{3}$,~\cite{2017LIU}, YH$_{4}$,~\cite{2017LIU} and YH$_{6}$.~\cite{2017LIU}
    }
    \label{fig.enthalpy_compare}
  \end{center}
\end{figure*}

\begin{figure*}[htbp]
  \begin{center}
    \includegraphics[width=\linewidth]{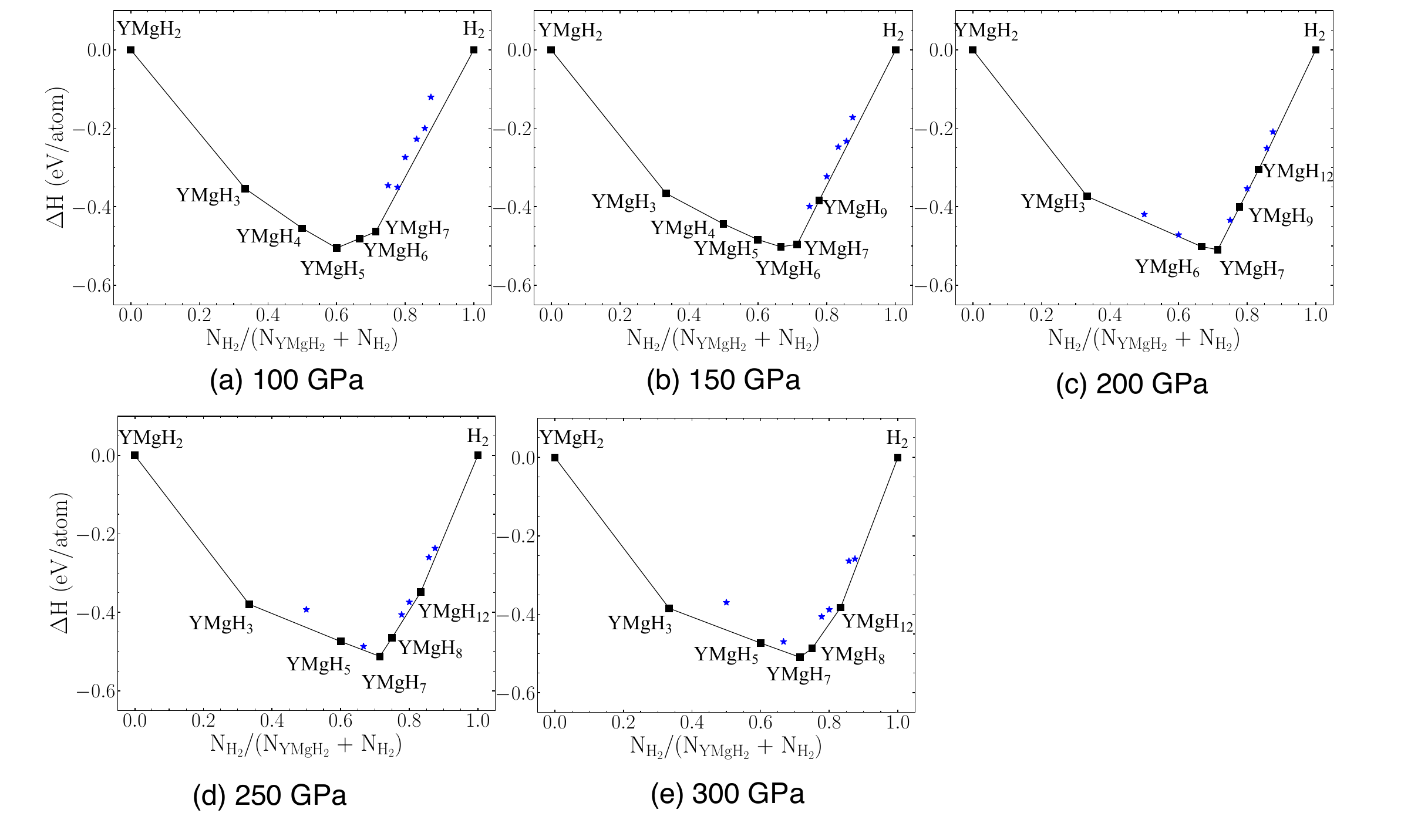}
    \caption{
Convex hull of YMgH$_x$ relative to YMgH$_{2}$  and H-$C2/c$~\cite{2007PIC} at the pressure of (a) 100 GPa, (b) 150 GPa, (c) 200GPa, (d) 250 GPa, and (e) 300 GPa.
    }
    \label{fig.cvh}
  \end{center}
\end{figure*}

\begin{figure*}[h]
  \begin{center}
    \includegraphics[width=\linewidth]{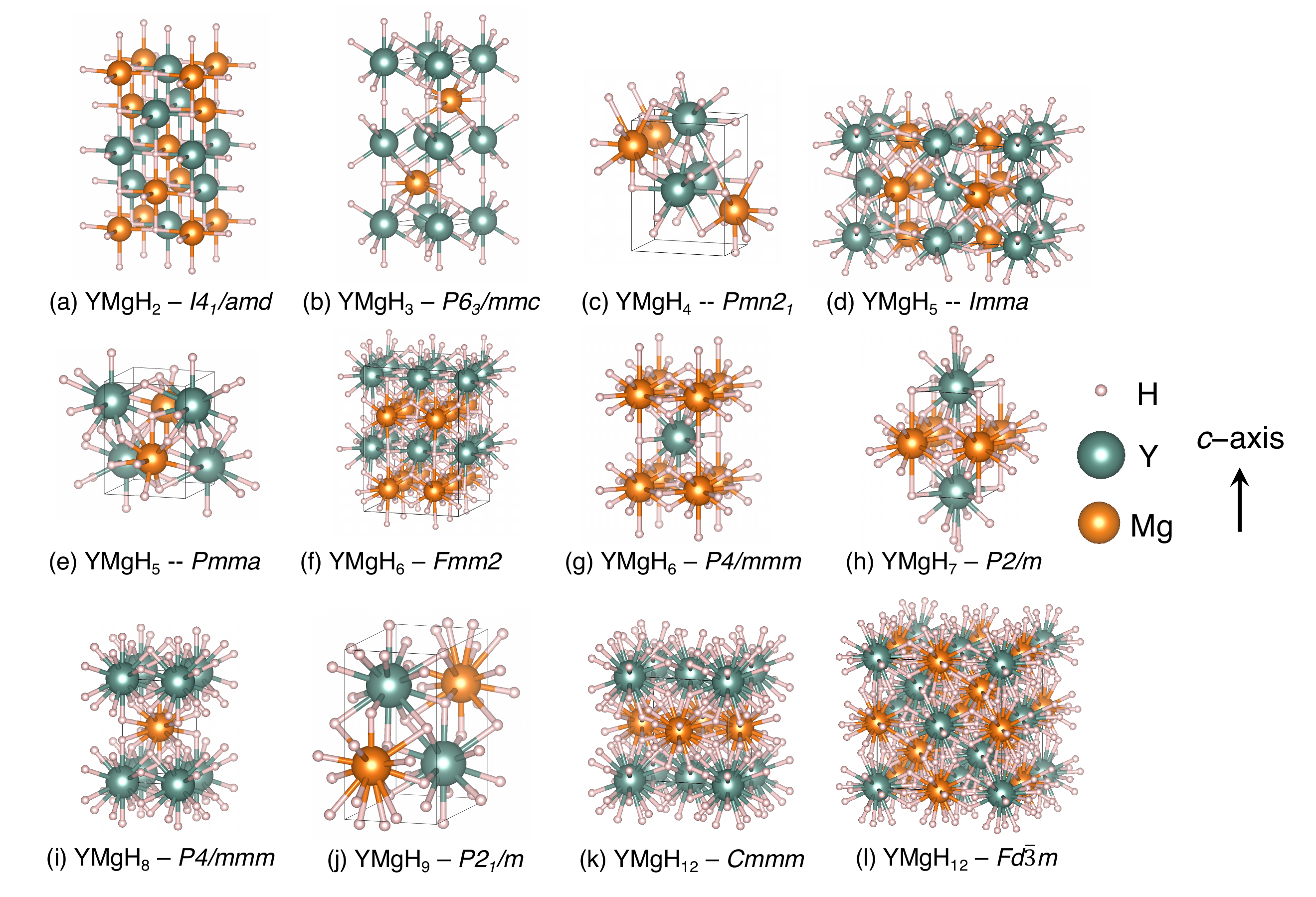}
    \caption{
Crystal structures of the predicted YMgH$_{x}$ phases.
    }
    \label{fig.crystal_structure}
  \end{center}
\end{figure*}
\begin{figure*}[h]
  \begin{center}
    \includegraphics[width=\linewidth]{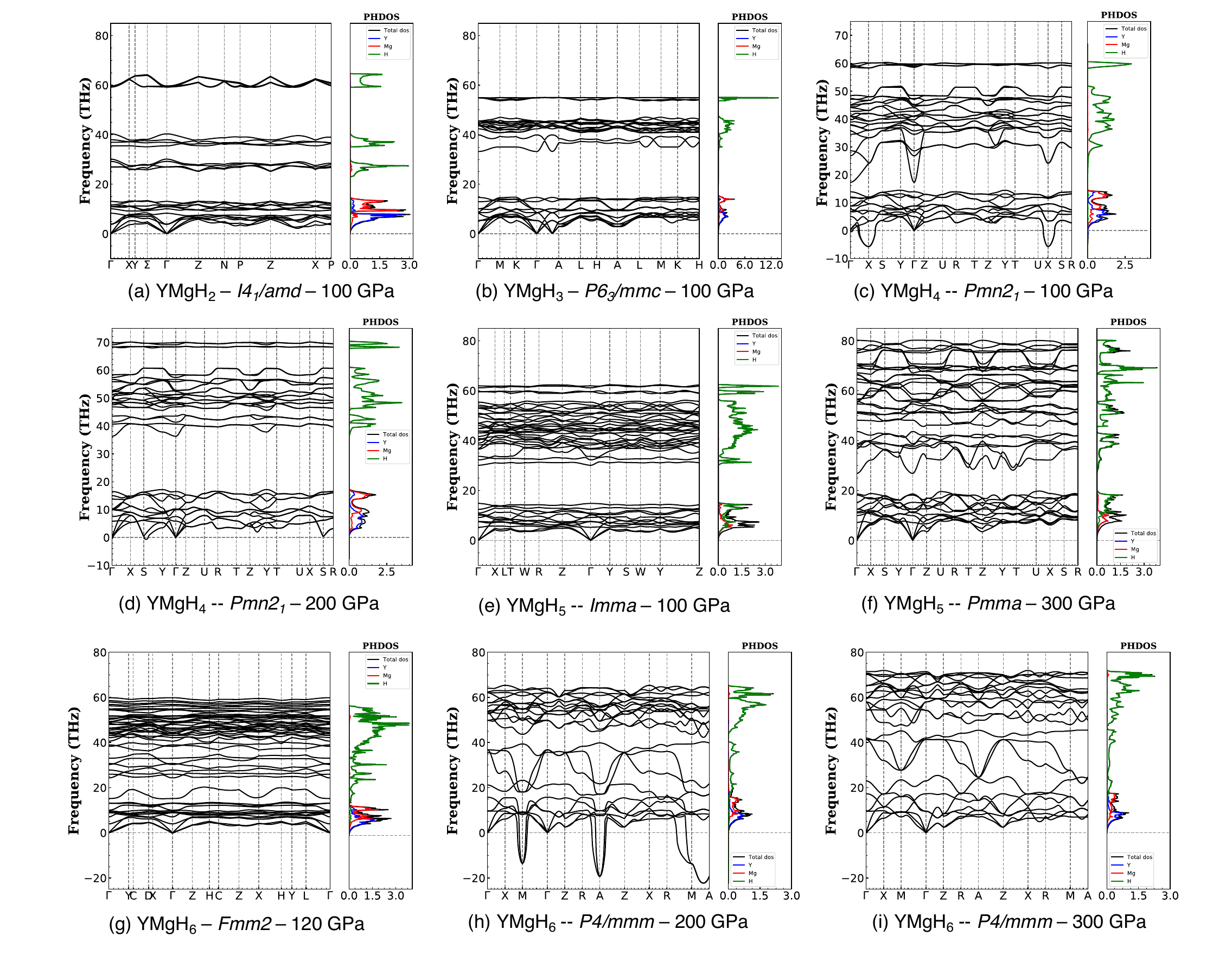}
    \caption{Phonon band structures and atom-projected phonon density of states of (a) YMgH$_2$--$I4_1/amd$ at 100 GPa, (b) YMgH$_3$--$P6_3/mmmc$ at 100 GPa, (c) YMgH$_4$--$Pmn2_1$ at 100 GPa, (d) YMgH$_4$--$Pmn2_1$ at 200 GPa, (e) YMgH$_5$--$Imma$ at 100 GPa, (f) YMgH$_5$--$Pmma$ at 300 GPa, (g) YMgH$_6$--$Fmm2$ at 120 GPa, (h) YMgH$_6$--$P4/mmm$ at 200 GPa, and (i) YMgH$_6$--$P4/mmm$ at 300 GPa.
    }
    \label{fig.phonon_1}
  \end{center}
\end{figure*}
\begin{figure*}[h]
  \begin{center}
    \includegraphics[width=\linewidth]{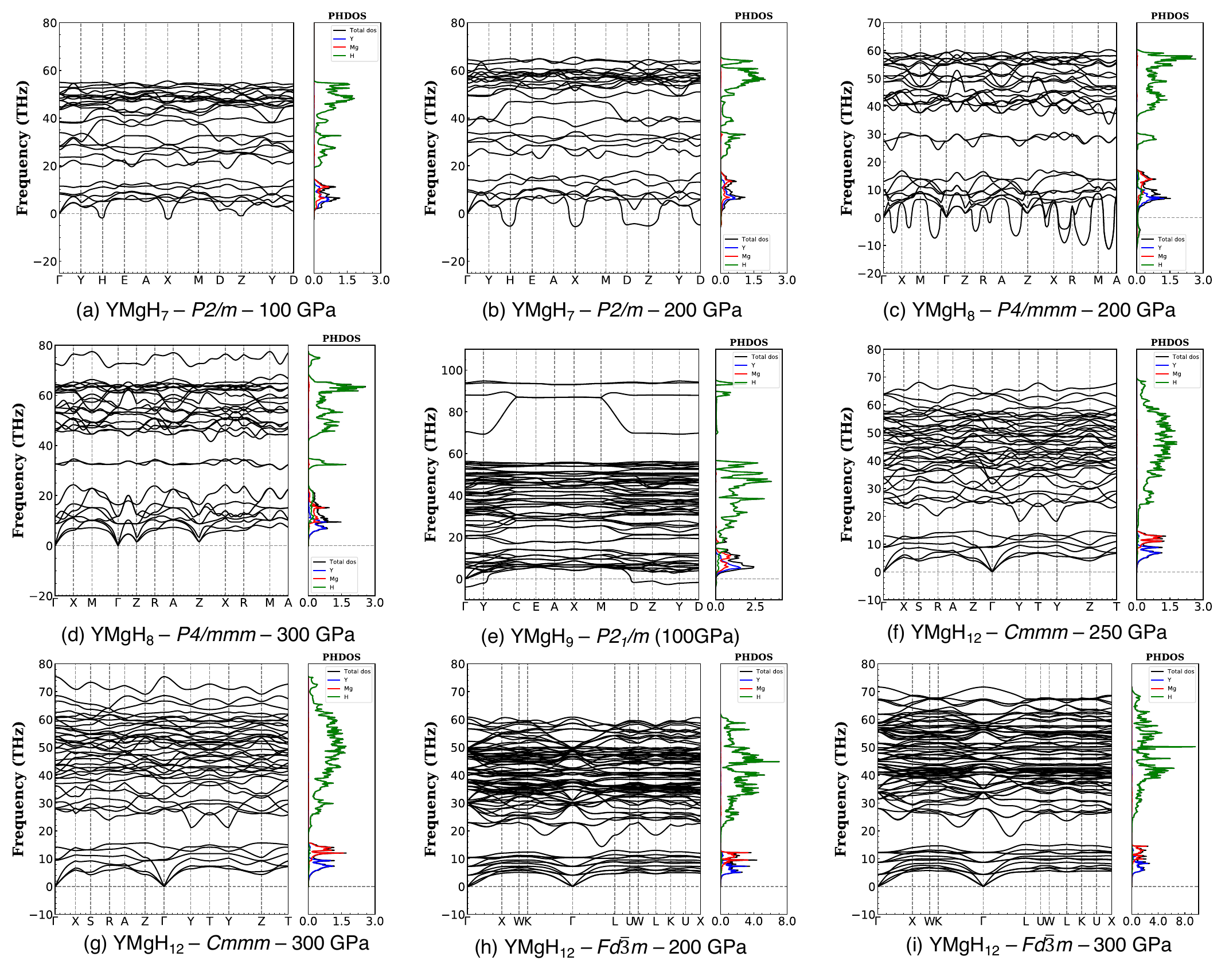}
    \caption{
Phonon band structures and atom-projected phonon density of states of (a) YMgH$_7$--$P2/m$ at 100 GPa, (b) YMgH$_7$--$P2/m$ at 200 GPa, (c) YMgH$_8$--$P4/mmm$ at 200 GPa, (d) YMgH$_8$--$P4/mmm$ at 300 GPa, (e) YMgH$_9$--$P2_1/m$ at 100 GPa, (f) YMgH$_{12}$--$Cmmm$ at 250 GPa, (g) YMgH$_{12}$--$Cmmm$ at 300 GPa, (h) YMgH$_{12}$--$Fd\bar{3}m$ at 200 GPa, and (i) YMgH$_{12}$--$Fd\bar{3}m$ at 300 GPa.
    }
    \label{fig.phonon_2}
  \end{center}
\end{figure*}

\begin{figure*}[h]
  \begin{center}
    \includegraphics[width=\linewidth]{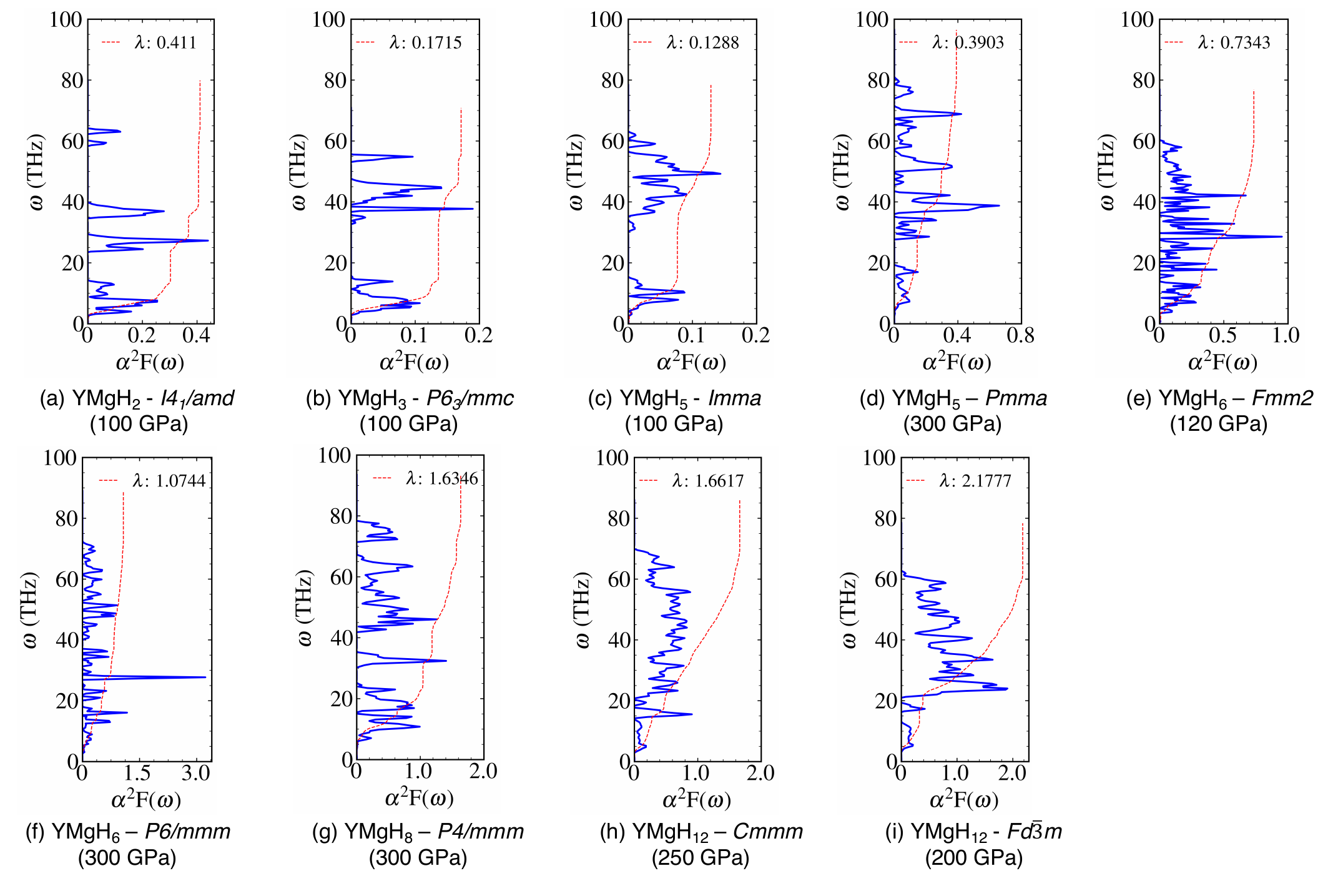}
    \caption{
Eliashberg spectral functions of (a) YMgH$_2$--$I4_1/amd$ at 100 GPa, (b) YMgH$_3$--$P6_3/mmc$ at 100 GPa, (c) YMgH$_5$--$Imma$ at 100 GPa, (d) YMgH$_5$--$Pmma$ at 300 GPa, (e) YMgH$_6$--$Fmm2$ at 120 GPa, (f) YMgH$_{6}$--$P6/mmm$ at 300 GPa, (g) YMgH$_{8}$--$P4/mmm$ at 300 GPa, (h) YMgH$_{12}$--$Cmmm$ at 250 GPa, and (i) YMgH$_{12}$--$Fd\bar{3}m$ at 200 GPa.
    }
    \label{fig.elph}
  \end{center}
\end{figure*}

\begin{figure*}[h]
  \begin{center}
    \includegraphics[width=\linewidth]{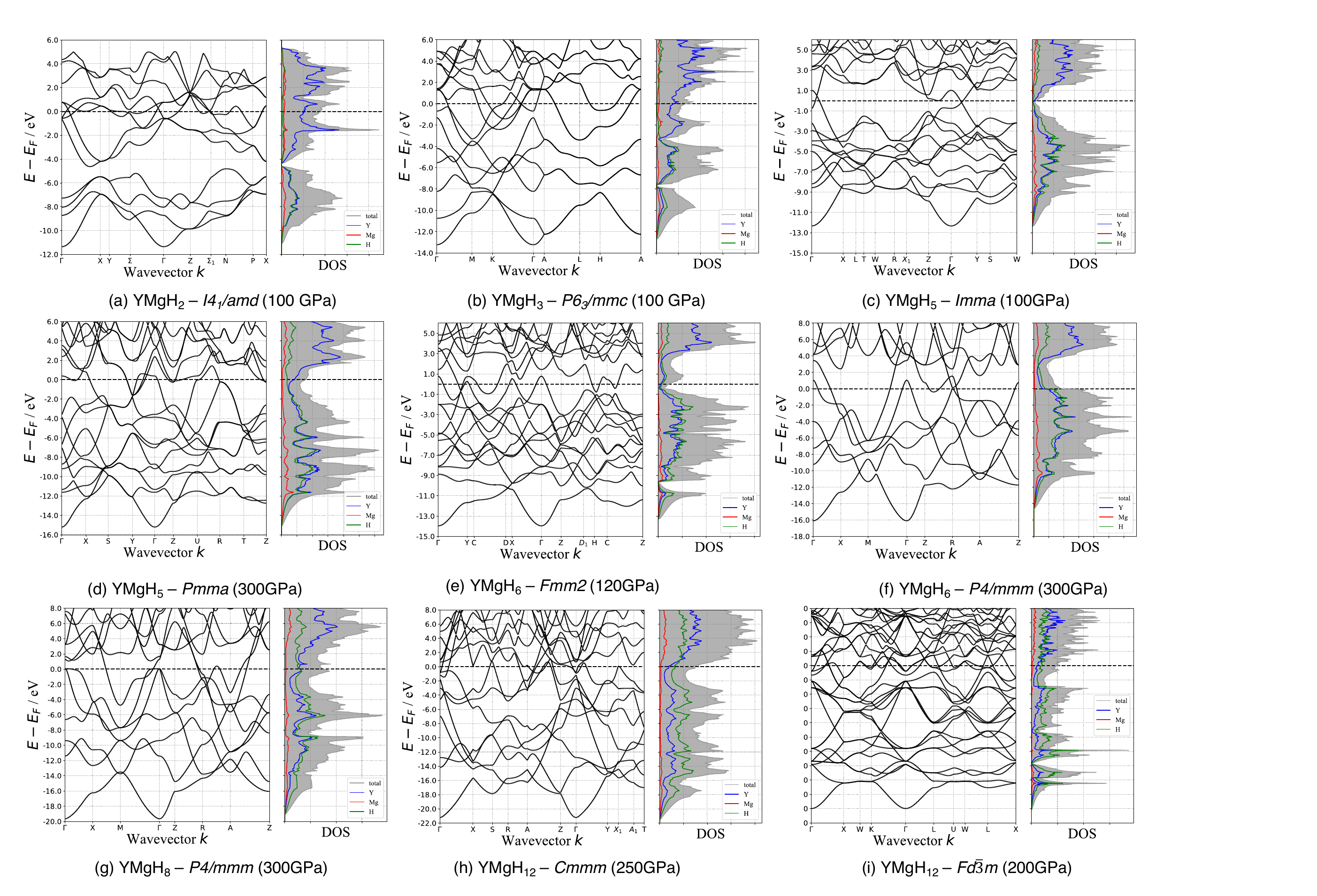}
    \caption{
Band structure and Density of state of the predicted YMgH$_{x}$ phases.
    }
    \label{fig.band}
  \end{center}
\end{figure*}
\begin{figure*}[h]
  \begin{center}
    \includegraphics[width=\linewidth]{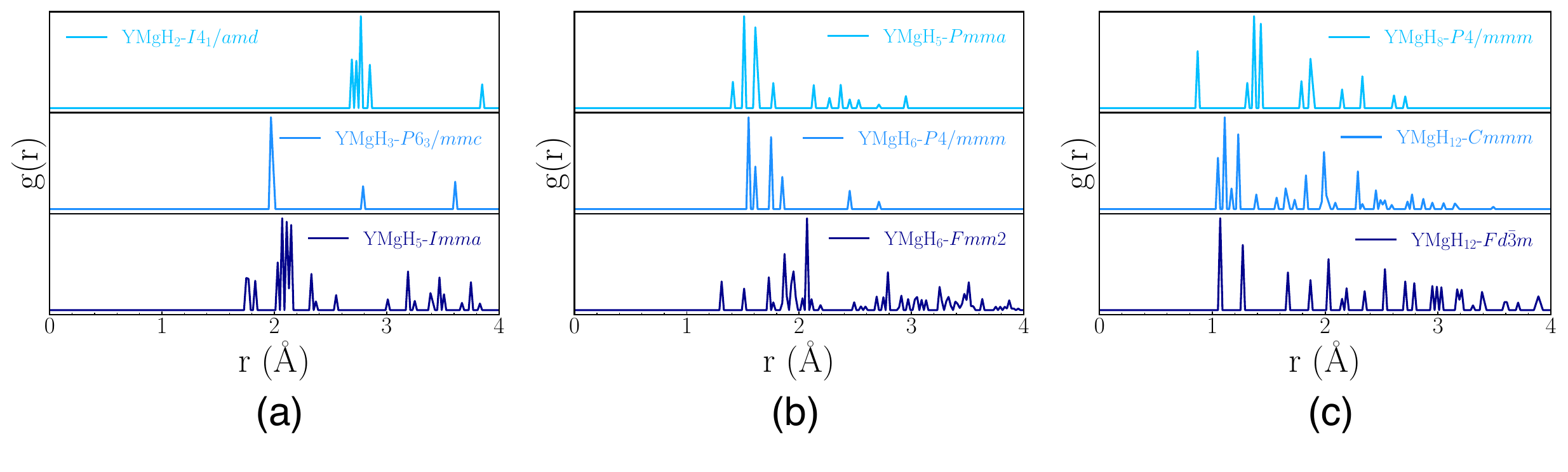}
    \caption{
The radial distribution function of the H-H atom pair in YMgH$_x$.
    }
    \label{fig.rdf}
  \end{center}
\end{figure*}
\begin{figure*}[h]
  \begin{center}
    \includegraphics[width=\linewidth]{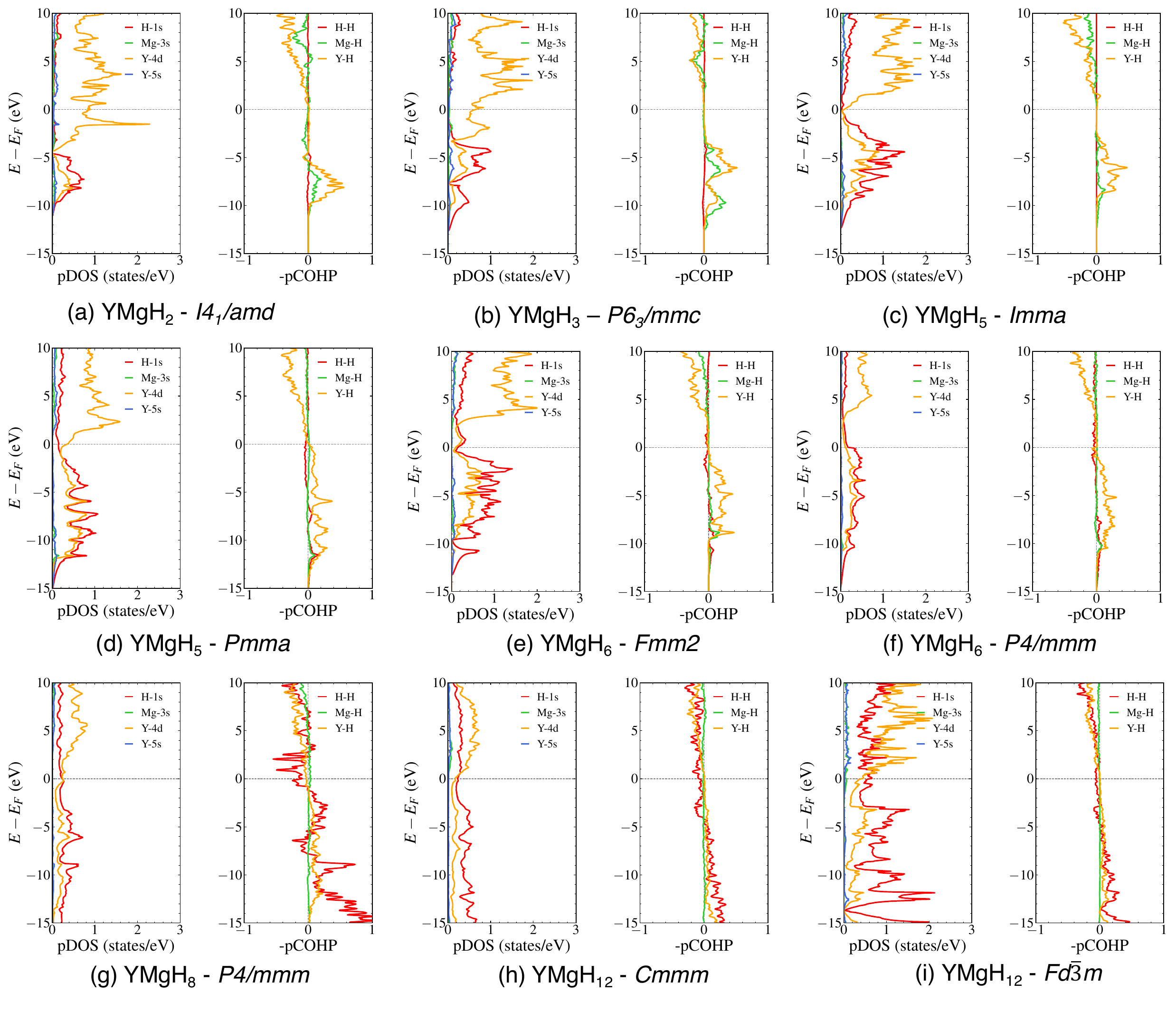}
    \caption{
The partial density of states (PDOS) and projected crystal orbital Hamilton population (pCOHP) of YMgH$_x$.
    }
    \label{fig.pdos_1}
  \end{center}
\end{figure*}

\clearpage
\newpage
\subsection{ Supplementary tables}

\begin{table*}[h]
 \begin{center}
   \caption{
 The distances of the first nearest neighboring Y-H, Mg-H, and H-H atom pairs.
   }
     \label{table.ICOHP}
\begin{tabular}{cccccccccccc}
\hline
\multicolumn{2}{l}{}             & YMgH$_{2}$       & YMgH$_{3}$               & YMgH$_{5}$       & YMgH$_{5}$       & YMgH$_{6}$       & YMgH$_{6}$      & YMgH$_{8}$      & YMgH$_{12}$      & YMgH$_{12}$     \\
\hline
   \specialrule{0em}{1pt}{1pt}
\multicolumn{2}{l}{Space group} & $I4_{1}/amd$   &$P6_{3}/mmc$                   & $Imma$        & $Pmma$        &$Fmm2$         & $P4/mmm$        & $P4/mmm$        & $Cmmm$        &  $Fd\bar{3}m$       \\
\hline
\multicolumn{2}{l}{Pressure (GPa)} &100         &100               &100         &300         &120      &300         &300         &250         &200         \\
\hline
\multirow{3}{*}{Distance ($\AA$)} & Y-H  & 1.931 & 2.028   & 1.904 & 1.728 & 1.899 & 1.758 & 1.821 & 1.879 &1.965\\
                          & Mg-H & 1.922        &  1.7035               &1.652         & 1.562        & 1.677        & 1.540        &  1.584       &1.726         &1.834         \\
                          & H-H  &2.681         & 1.971             &1.778         & 1.517        &  1.725       & 1.540        & 0.864        & 1.040        & 1.097        \\
 \hline
\end{tabular}
 \end{center}
\end{table*}
\begin{table*}[h]
 \begin{center}
   \caption{
     $T_c$ estimated by McMillan formula using
     {\it ab initio} phonon calculations
     for
 	YMgH$_{x}$
     at each pressure.
     $\lambda$ and $\omega_{\rm log}$ are
     the parameters appearing in the formula.
   }
     \label{table.Pdep}
\begin{tabular}{rrccrcc}
\hline
   \specialrule{0em}{1pt}{1pt}
Compound &Space group & $P$~(GPa) & $\lambda$ & $\omega_{\rm log}$~(K) & $N_{E_\mathrm{F}}$ (states/eV/ $\AA^{3}$) &$T_\mathrm{c}$~(K)  \\
 &   &           &   &    & &at $\mu$ = 0.1 -- 0.13\\
 \hline
YMgH$_{2}$ & $I4_{1}/amd$  & 100        & 0.411    & 451.20 & 0.0292 &  2.20 -- 1.01  \\
 \hline
YMgH$_{3}$ & $P6_{3}/mmc$  & 100        & 0.1715   & 469.93 & 0.0200 &  $\ll$ 0.001  \\
\hline
YMgH$_{5}$ & $Imma$  & 100        & 0.1288   & 761.1124 & 0.0046 &  $\ll$ 0.001  \\
\hline
YMgH$_{5}$ & $Pmma$ & 200     & 0.4247  & 1036.858 &   0.0121 & 6.02 -- 2.92 \\
 &  & 300     & 0.3903  & 1219.4306 &   0.0141  & 4.44 -- 1.83 \\
 \hline
  YMgH$_{6}$ & $Fmm2$ & 120 & 0.7343           &800.4893  & 0.0093 &  31.17 -- 24.39  \\
\hline
YMgH$_{6}$ & $P4/mmm$  & 300        & 1.0744    & 963.2549  & 0.0210 & 74.55 -- 64.91     \\
\hline
YMgH$_{8}$ & $P4/mmm$  & 300        & 1.6346    & 1012.8135  & 0.0249&  124.77 -- 114.70    \\
\hline
YMgH$_{12}$ & $Cmmm$  & 250         & 1.6617    & 1224.0843  & 0.0209 &  152.92 -- 140.78    \\
 &   & 300         & 1.3054    &1449.2025 & 0.0211 & 143.19 -- 128.52   \\
\hline
   \specialrule{0em}{1pt}{1pt}
YMgH$_{12}$ & $Fd\bar{3}m$  & 200        & 2.1777   & 1249.5972   & 0.0248 &  190.01 -- 178.22    \\
 &   & 300        & 1.8107    & 1408.7617   & 0.0271 &  188.55 -- 174.78    \\
\hline
\end{tabular}
 \end{center}
\end{table*}

\begin{center}
\begin{longtable}[htbp]{c c c c lccc}
\caption[aaaa]{Crystal structures of Y-Mg-H predicted at each pressure~($P$). Lattice parameters ($a$, $b$ and $c$) are given in unit of $\AA$.}\\ \hline
\endfirsthead
\multicolumn{8}{c}%
{{\bfseries \tablename\ \thetable{} -- continued from previous page}} \\
\hline \multicolumn{1}{c}{Compound} &
\multicolumn{1}{c}{Space group} &
\multicolumn{1}{c}{$P$~(GPa)}&
\multicolumn{1}{c}{Lattice parameters} &
\multicolumn{4}{c}{Atomic coordinates (fractional)}\\
&&&&   \multicolumn{1}{c}{Atoms} &\multicolumn{1}{c}{$x$}& \multicolumn{1}{c}{$y$} &\multicolumn{1}{c}{$z$}\\
\hline
\endhead
\hline \multicolumn{8}{r}{{Continued on next page}} \\ \hline
\endfoot
\hline \hline
\endlastfoot
Compound & Space group & $P$~(GPa)& Lattice parameters & \multicolumn{4}{c}{Atomic coordinates (fractional)}\\
&&&&    Atoms  & $x$ &  $y$  & $z$ \\
\hline
\specialrule{0em}{1pt}{1pt}
YMgH$_{2}$    & $I4_{1}/amd$ & 100 &  $a = b =  3.8598$ &   Y(4$b$)&  0.00000 &  0.00000 & 0.50000 \\
              &              &     &  $c = 7.9269$      &   Mg(4$a$) &0.00000 &  0.00000 & 0.00000 \\
              &              &     &  $\alpha = \beta = \gamma =  90^{\circ}$      &  H(8$e$)  &0.00000 &  0.00000 & 0.24241  \\
 \hline
   \specialrule{0em}{1pt}{1pt} 
YMgH$_{3}$    & $P6_{1}/mmc$ & 100 &   $a = b =  2.95056$  &  Y(2$a$) &  0.00000  & 0.00000 &0.00000 \\
              &              &      &  $c = 8.36345$ &       Mg(2$d$) &  0.33333&  0.66667 &  0.75000 \\
              &              &      & $\alpha = \beta =  90^{\circ}$  &  H(4$f$) & 0.33333 & 0.66667 & 0.36848\\
              &              &      & $\gamma =  120^{\circ}$  & H(2$b$) & 0.00000  & 0.00000 &0.25000\\
\hline
   \specialrule{0em}{1pt}{1pt}
YMgH$_{4}$    & $Pmn2_{1}$ & 100 &  $a = 3.0249$ &  Y(2$a$) & 0.00000 & 0.50190 & 0.48731 \\
              && &  $b = 4.0812$  &Mg(2$a$) & 0.00000 &0.01506 &0.81030\\
              &&&$c = 5.5122$&  H(2$a$) &0.00000 &0.00618 &0.48454\\
              &&& $\alpha = \beta = \gamma =  90^{\circ}$ & H(2$a$)&0.00000&0.23584& 0.14577\\
              &&&  &H(2$a$) &0.00000&0.59877& 0.81387\\
              &&&  &H(2$a$) &0.00000&0.75436& 0.16581\\
\hline
   \specialrule{0em}{1pt}{1pt}
YMgH$_{5}$ & $Imma$ & 100  & $a = 3.5536$ &  Y(4$a$)  & 0.00000  &0.00000  &0.00000 \\
  &&&$b = 8.3654$ & Mg(4$e$) & 0.00000  &0.25000  &0.50640 \\
 &&&$c = 4.9179$  &  H(8$h$)  & 0.00000  &0.07060  &0.36745 \\
 &&&$\alpha = \beta = \gamma = 90^{\circ}$&H(8$g$)  & 0.25000  &0.14083  &0.75000 \\
&&&  &H(4$e$)  & 0.00000  &0.25000  &0.09003 \\
\hline
   \specialrule{0em}{1pt}{1pt}
YMgH$_{5}$& $Pmma$ & 200 &$a = 3.6996$  &    Y(2$e$)  &0.25000  &0.00000 & 0.29815  \\ 
 &&&$b = 3.4314$ &Mg(2$f$) &0.25000  &0.50000 & 0.80204 \\
 &&&$c = 4.5078$ &H(4$g$)  &0.00000  &0.21208 & 0.00000 \\
 &&&$\alpha = \beta = \gamma = 90^{\circ}$ &H(2$d$)  &0.00000  &0.50000 & 0.50000 \\
&&&&H(2$e$)  &0.25000  &0.00000 & 0.72445 \\
&&&&H(2$f$)  &0.25000  &0.50000 & 0.18862 \\
\hline
   \specialrule{0em}{1pt}{1pt}
YMgH$_{6}$ & $Fmm2$ & 100 &$a = 5.8175$ &Y(8$d$)  & 0.24964  &  0.00000 &0.41497\\  
 &&&$b = 5.9495$ &Mg(8$c$) & 0.00000  &  0.24126 &0.66623  \\
 &&&$c = 8.7087$& H(8$c$)  & 0.00000  &  0.24353 &0.92130  \\
 &&&$\alpha = \beta = \gamma = 90^{\circ}$ &H(8$d$)  & 0.18399  &  0.00000 &0.14984  \\
&&&&H(8$b$)  & 0.25000  &  0.25000 &0.05659  \\
&&&&H(8$b$)  & 0.25000  &  0.25000 &0.27322  \\
&&&&H(4$a$)  & 0.00000  &  0.00000 &0.06899  \\
&&&&H(4$a$)  & 0.00000  &  0.00000 &0.26886  \\
&&&&H(4$a$)  & 0.00000  &  0.00000 &0.55987  \\
&&&&H(4$a$)  & 0.00000  &  0.00000 &0.77059  \\
\hline
   \specialrule{0em}{1pt}{1pt}
YMgH$_{6}$ & $P4/mmm$ & 300 & $a = 2.6235$ &Y(1$b$)    &0.00000  & 0.00000  &0.50000 \\
 &&&$b = 2.6235$ &Mg(1$c$)   &0.50000  & 0.50000  &0.00000 \\
 &&&$c = 3.9533$ &H(4$i$)    &0.00000  & 0.50000  &0.20410 \\
 &&&$\alpha = \beta = \gamma = 90^{\circ}$  &H(1$a$)    &0.00000  & 0.00000  &0.00000 \\
&&& &H(1$d$)    &0.50000  & 0.50000  &0.50000 \\
\hline
   \specialrule{0em}{1pt}{1pt}
YMgH$_{7}$ & $P2/m$ & 150 & $a = 2.8255$ &Y(1$e$)  & 0.50000 & 0.50000  & 0.00000  \\
 &&&$b = 2.8664$ & Mg(1$c$) & 0.00000 & 0.00000  & 0.50000  \\
 &&&$c = 4.3458$& H(2$n$)  & 0.00699 & 0.50000  & 0.28870  \\
 &&&$\alpha = \gamma = 90^{\circ}$ &H(2$n$)  & 0.38478 & 0.50000  & 0.44923  \\
 &&&$\beta = 90.8145^{\circ}$  &H(2$m$)  & 0.49818 & 0.00000  & 0.28506  \\
&&& &H(1$a$)  & 0.00000 & 0.00000  & 0.00000  \\
\hline
   \specialrule{0em}{1pt}{1pt}
YMgH$_{8}$ & $P4/mmm$ & 300 & $a = 2.5271$  &Y(1$a$)  &0.00000 & 0.00000 & 0.00000 \\
 &&&$b = 2.5271$ &Mg(1$d$) &0.50000 & 0.50000 & 0.50000 \\
 &&&$c = 4.5369$ &H(4$i$)  &0.00000 & 0.50000 & 0.28857 \\
 &&&$\alpha = \beta = \gamma = 90^{\circ}$ &H(2$g$)  &0.00000 & 0.00000 & 0.40187 \\
&&&&H(2$h$)  &0.50000 & 0.50000 & 0.14891 \\
\hline
   \specialrule{0em}{1pt}{1pt}
YMgH$_{9}$ & $P2_{1}/m$ & 100 & $a = 5.1702$  &Y(2$e$) &0.41530&0.75000  & 0.26439\\
 &&&$b = 3.0745$ &Mg(2$e$)&0.07454&0.75000  & 0.68683\\
 &&&$c = 5.4932$ & H(4$f$) &0.08755&0.61229  & 0.03743\\
 &&&$\alpha =  \gamma = 90^{\circ}$ & H(2$e$) &0.06687&0.75000  & 0.37652\\
 &&&$\beta = 90.1008^{\circ}$ &H(2$e$) &0.20436&0.25000  & 0.79435\\
&&&&H(2$e$) &0.20888&0.25000  & 0.14128\\
&&&&H(2$e$) &0.24539&0.25000  & 0.93789\\
&&&&H(2$e$) &0.26150&0.25000  & 0.48192\\
&&&&H(2$e$) &0.38701&0.75000  & 0.61655\\
&&&&H(2$e$) &0.42098&0.75000  & 0.91256\\
\hline
   \specialrule{0em}{1pt}{1pt}
YMgH$_{12}$ & $Cmmm$ & 250 &$a = 3.3296$ &Y(2$a$)  &0.00000 & 0.00000   & 0.22450      \\
 &&&$b = 4.6364$  &Mg(2$c$) &0.50000 & 0.50000   & 0.74629      \\
 &&&$c = 4.6931$ & H(8$n$)  &0.12525 & 0.38836   & 0.46001      \\
 &&&$\alpha = \beta = \gamma = 90^{\circ}$ & H(8$n$)  &0.38787 & 0.14919   & 0.52716      \\
&&&&H(8$m$)  &0.25000 & 0.28656   & 0.03551      \\
\hline
   \specialrule{0em}{1pt}{1pt}
YMgH$_{12}$ & $Fd\bar{3}m$ & 250 & $a = b = c =  6.6105$ &Y(8$a$) & 0.00000 & 0.00000&0.50000  \\
 &&&$\alpha = \beta = \gamma = 90^{\circ}$ &Mg(8$b$) &0.00000 & 0.00000&0.00000  \\
&&&&H(96$h$) & 0.00934 & 0.12500&0.24066  \\
\hline
\end{longtable}
\end{center}

\clearpage
\bibliographystyle{apsrev4-1}
\bibliography{references}